\documentstyle[12pt,epsfig]{iopart}

\input axodraw.sty
\def\np#1#2#3   {{ Nucl. Phys.} {\bf#1}, #2 (#3) }
\def\pcps#1#2#3 {{ Proc. Cam. Phil. Soc.} {\bf#1}, #2 (#3) }
\def\pl#1#2#3   {{ Phys. Lett.} {\bf#1}, #2 (#3) }
\def\prep#1#2#3 {{ Phys. Rep.} {\bf#1}, #2 (#3) }
\def\prev#1#2#3 {{ Phys. Rev.} {\bf#1}, #2 (#3) }
\def\prl#1#2#3  {{ Phys. Rev. Lett.} {\bf#1}, #2 (#3) }
\def\prs#1#2#3  {{ Proc. Roy. Soc.} {\bf#1}, #2 (#3) }
\def\ptp#1#2#3  {{ Prog. Th. Phys.} {\bf#1}, #2 (#3) }
\def\rmp#1#2#3  {{ Rev. Mod. Phys.} {\bf#1}, #2 (#3) }
\def\rpp#1#2#3  {{ Rep. Prog. Phys.} {\bf#1}, #2 (#3) }
\def\zp#1#2#3   {{ Z. Phys.} {\bf#1}, #2 (#3) }
\def\epj#1#2#3   {{ Eur. Phys. Jour.} {\bf#1}, #2 (#3) }

\newcommand{\geqsim}{\,\raisebox{-0.6ex}{$\buildrel > \over \sim$}\,}

\newcommand{\Ftwo}   {\mbox{$F_2$}}
\newcommand{\QQ}{\mbox{${Q^2}$}}

\begin{document}

\pagestyle{empty}
\rightline{OUNP-99-04}
\rightline{MZ-TH-99-02}
\title
{High-$Q^2$ Deep Inelastic Scattering at HERA}

\author{A M Cooper-Sarkar}
\address{Department of Physics, Univeristy of Oxford,
Oxford OX1 3RH, UK}

\author{ A Bodek}

\address{Department of Physics and Astronomy,
University of Rochester, Rochester, NY 14627 }

\author{K  Long}
\address{Imperial College of Science and Technology, 
London SW7 2BZ,UK}
\author{E Rizvi}

\address{School of Physics, Birmingham University, 
Birmingham B15 2TT, UK}

\author{ H Spiesberger} 

\address{Institut f\"ur Physik, Johannes-Gutenberg-Universit\"at
  D-55099 Mainz, Germany}

\begin{abstract}
High $Q^2$ NC and CC cross sections as measured at HERA can give information
on two distinct areas of current interest. Firstly, supposing that all the
electroweak parameters are well known, these cross-sections may be used to
give information on parton distribution functions (PDFs) in the new kinematic 
regime at high $x$ and high $Q^2$. Secondly, supposing that PDFs are well known
after evolution in $Q^2$ from the kinematic regime where they are already
measured, these cross-sections may be used to give information on
electroweak parameters in a process where the exchanged boson is `spacelike' 
rather than `timelike'. WG1
addressed itself to clarifying the limits of our present and possible future
knowledge on both these points.

\end{abstract}

\nopagebreak



\section{Introduction}

The differential (Born) cross-section for charged 
lepton-nucleon scattering, mediated by the neutral current at high $Q^2$ and 
accounting for the possible polarization of the lepton beam, is given by
\begin{equation}
 \frac {d^2\sigma (l^{\pm}N) } {dxdQ^2} =  \frac {2\pi\alpha^2} {Q^4 x}  
\left[Y_+\,F_2^{lN}(x,Q^2) \mp Y_-\, xF_3^{lN}(x,Q^2) \right],
\label{eq:NCxsec}
\end{equation} 
where $F_2$ is expressed in terms of parton distributions as
\begin{equation}
 F_2^{lN}(x,Q^2) = \Sigma_i A_i^{L,R}(Q^2)\times (xq_i(x,Q^2) + x\bar q_i (x,Q^2)),
\label{eq:f2ln}
\end{equation}
and the sum is taken over flavours of quarks which are above threshold.
(A contribution from the longitudinal structure function, $F_L$, is
neglected in Eq.~\ref{eq:NCxsec}).
$A_i$ gives the couplings of the quarks and leptons to the currents
\begin{equation}
        A_i^{L,R}(Q^2) = e_i^2 + 2e_i e_l(v_l \pm a_l)v_i P_Z + (v_l \pm a_l)^2
(v_i^2 +a_i^2) P_Z^2 
\label{eq:ai}
\end{equation}
These couplings depend on whether the
polarization of the lepton beam is left ($L$, upper signs) or right 
($R$, lower signs) handed. The
notation $e_l$ specifies the incoming lepton's charge such that 
$e_l= -1$. The 
vector and axial-vector couplings of the fermions are given by
\begin{equation}
 v_f = (T_{3f} - 2 e_f \sin^2\theta_W),\ \ a_f = T_{3f}
\end{equation}
where the definition holds good for any fermion, whether lepton or quark;
$T_{3f}$ is the weak isospin, and $\theta_W$ is the Weinberg 
angle~\footnote[1]{Neutrinos and charged leptons of the same 
family form weak 
isospin doublets with $T_3 = 1/2,-1/2$ respectively; and the  
quarks form similar weak isospin doublets, within the families 
$(u,d),(c,s),(t,b)$, with $T_3 = 1/2, -1/2$ respectively. For antiparticles
both $T_{3f}$ and $e_f$ change sign.}.
The term  $P_Z$ accounts for the $Z$ propagator
\begin{equation}
           P_Z = \frac {Q^2} {Q^2 + M_Z^2} \frac{\sqrt{2}G_{\mu}M_Z^2}
       {4\pi\alpha}
\end{equation}
Thus one can identify the contributions of $\gamma$ exhange, $\gamma-Z$ 
interference and $Z$ exchange in the coupling $A_i$.
Finally the parity violating structure function $xF_3$ is given by
\begin{equation}
  xF_3^{lN}(x,Q^2) = \Sigma_i B_i^{L,R}(Q^2)\times (xq_i(x,Q^2) - x\bar q_i (x,Q^2)),
\label{eq:xf3ln}
\end{equation}
where
\begin{equation}
        B_i^{L,R}(Q^2) =  \mp 2e_i e_l(v_l \pm a_l)a_i P_Z \pm 2(v_l \pm a_l)^2
v_i a_i P_Z^2.
\label{eq:bi}
\end{equation}
\vspace{3mm}
The corresponding cross-sections for antilepton scattering are given by
swapping $L \rightarrow R$, $R\rightarrow L$ in the expressions for $F_2$ and
$xF_3$ given in Eq.~\ref{eq:f2ln} and Eq.~\ref{eq:xf3ln}, and by substituting
the antilepton charge $e_l = +1$.

The differential (Born) cross-sections for charged lepton-nucleon scattering,
mediated by the charged current are given by
\begin{equation}
\frac {d^2\sigma^{CC}(l^\pm N) } {dxdQ^2} =  \frac {G_{\mu}^2} {4\pi x} \frac {
 M^4_W} { (Q^2 + M_W^2)^2 }\left[  
 Y_+\,F_2(x,Q^2)  \mp Y_-\, xF_3(x,Q^2) \right].
\label{eq:CCxsec}
\end{equation}
Expressing the structure functions in terms of parton distributions and
accounting for polarization of the lepton beam gives
\begin{eqnarray}
\displaystyle \fl
\frac {d^2\sigma^{CC}(l^-N) } {dxdQ^2} = (1 - P)\frac {G_{\mu}^2} {2\pi x} \frac 
{ M^4_W} { (Q^2 + M_W^2)^2 }   \left[ \Sigma_{i=u,c}
 xq_i(x,Q^2) + (1-y)^2 \Sigma_{i=d,s} x\bar q_i (x,Q^2) \right]
\label{eq:empCC}
\end{eqnarray}
whereas for antilepton scattering we have
\begin{eqnarray}
\displaystyle \fl
\frac {d^2\sigma^{CC}(l^+N) } {dxdQ^2} = (1 + P)\frac {G_{\mu}^2} {2\pi x } \frac 
{ M^4_W} { (Q^2 + M_W^2)^2 }    \left[ 
(1-y)^2 \Sigma_{i=d,s} xq_i(x,Q^2) +  \Sigma_{i=u,c} x\bar q_i (x,Q^2) \right]
\label{eq:eppCC}
\end{eqnarray}
where the sums contain only the appropriate quarks or antiquarks for the 
charge of the current and the polarization of the lepton beam, 
$P = (N_R - N_L)/(N_R + N_L)$.

Clearly measurements of these cross-sections can give information both on
electroweak parameters and on parton distribution functions (PDFs).
In Sec.\ref{sec:eram} and Sec.\ref{sec:bodek} we consider how accurately we really know
PDFs based on lower $Q^2$ measurements, and hence how much HERA high $Q^2$ 
measurements will improve our knowledge.
Searches for new physics rely crucially on accurate determinations of
event rates or cross-sections from standard model (SM) processes. 
In lepton-hadron
or hadron-hadron collisions, this translates into a need to estimate 
uncertainties on the PDFs of the incident
hadron. An example of such a need was seen when in 1997 both HERA experiments
observed a slight excess of events at very high \QQ~\cite{highQ2}.
Uncertainties on the SM cross-sections have been obtained by varying the 
input PDFs in the Monte-Carlo models. This method may not reflect the true
PDF uncertainty, but rather highlight the different approaches used by
global fitters in their PDF extractions. More reliable estimates of the
uncertainties on PDFs are considered in Sec.~\ref{sec:eram}.

Recent work on parton distributions functions (PDF's)
in the nucleon has focussed
on probing the sea and gluon distribution at small $x$. The valence 
quark distributions have been considered to be relatively well understood.
However, this is not really true at $x \geqsim 0.5$.
In particular our knowledge of the $d$ quark distribution at high $x$ is
extracted from data taken on deuteron targets, and nuclear binding effects
in the deuteron have not always been properly accounted. Our knowledge of the
$u$ quark density at high $x$ has also been questioned by Kuhlmann et al, 
who recently proposed a toy model~\cite{toy} which included
 the possibility of an additional contribution to the $u$ quark distribution
(beyond $x>0.75$) as an explanation for both the initial HERA high
 $Q^2$ anomaly~\cite{highQ2}
and for the jet excess at high-$P_T$  at CDF~\cite{CDFjet}.
A precise knowledge of the $u$ and $d$ quark distribution
at high $x$ is very important at collider energies in order to 
estimate backgrounds to signals
for new physics accurately.
In addition, the value of $d/u$ as $x \rightarrow 1$ is of theoretical
interest in its own right. It is clear from Eqs.~\ref{eq:empCC} 
and~\ref{eq:eppCC} that HERA CC data can greatly improve our knowledge of
these high $x$ valence quark densities. 
We discuss the current status of our knowledge 
in Sec.~\ref{sec:bodek}.

In Sec.~\ref{sec:hubert} and Sec.~\ref{sec:ken} we discuss the possibilty
of using high $Q^2$ HERA data to gain information on electroweak parameters.
We have specified the Born cross-sections for the processes above but
if we aim for an eventual precision of $1\,\%$ then electroweak
radiative corrections of order $O(\alpha)$ have to be taken into
account. We discuss the current status of implementation of the full
electroweak radiative corrections and the implications for fitting
electroweak parameters in Sec.~\ref{sec:hubert}. 

Measurements from LEP and the Tevatron tightly constrain the electroweak
sector of the Standard Model (SM).
In particular, the mass of the $W$-boson, $M_W$, is known precisely 
from measurements of the properties of real, `time-like', $W$-bosons.
In the SM description of charged current (CC) deep inelastic
scattering (DIS) this same mass appears in the `propagator' of the
exchanged `space-like' $W$-boson (see Eq.~\ref{eq:CCxsec}). 
Hence, for the SM to be self-consistent the same value of
$M_W$ must be obtained from measurements of the space-like `propagator
mass' in CC DIS as is obtained from LEP and the Tevatron; any
difference indicates new physics.
In Sec.~\ref{sec:ken} we discuss the present and 
future achievable accuracy of fits to $M_W$ and $G_\mu$ using HERA data.

The workshop necessarily concerned itself with results which can be 
achieved at present or in the near future.
In Sec.~\ref{sec:summary} we give a summary and discuss prospects for the
more distant future.

\section{Uncertainties on Parton Distributions}
\label{sec:eram}

\subsection{Experimental Uncertainties}
Measurements of cross-sections and structure functions inevitably
involve detailed estimates of systematic uncertainties which are
usually published
with the measurements. The systematic uncertainties may be classified into
two types: uncorrelated systematics in which the uncertainties fluctuate
point-to-point (for example limitations on the statistics generated in
Monte Carlo simulations of systematic effects) and correlated systematics 
which affect the data points in a 
coherent manner (for example, uncertainty in the luminosity determination). 
These systematic uncertainties may be propagated into the QCD fits used to 
extract parton densities and thereby allow an estimate to be made of 
experimental uncertainties in the extracted PDFs. A detailed
derivation of the formalism necessary to make such estimates 
is given in~\cite{zomer} and 
both the H1 and ZEUS collaborations have propagated systematic errors into 
their estimates of the errors on the gluon density
at low $x$~\cite{H194,zglu}. These experimental uncertainties are not the
focus of the present section. They are thoroughly
discussed by Botje~\cite{botje}.

\subsection{`Theoretical' Uncertainties}

In addition to experimental uncertainties there are uncertainties
which are of a more theoretical nature. 
A list of such sources of uncertainty is given below.

\begin{itemize}
\item {\boldmath $\alpha_s$.} The size of $\alpha_s$ governs the 
NLO \QQ~evolution
of the PDFs. The Particle Data Group estimates an uncertainty of $\pm 0.002$ 
at a scale of $M_Z^2$~\cite{pdg}.
\item {\bf DGLAP solution methods.} The DGLAP equations may be solved 
numerically 
using two classes of methods, namely $x-$space methods and 
$n-$space methods. Differences
between these two approaches are generally small, but can be as large as 7\% 
for the
gluon density at low $x$~\cite{heraproc}
\item {\bf Heavy Quark Treatment.} The way that heavy quarks are 
treated in  
DGLAP evolution can lead to uncertainties. In one approach the heavy 
quarks ($c$,$b$) 
are treated as massless partons and are evolved only 
above \QQ~thresholds which are usually 
set equal to the masses squared. This method is not expected to 
give an accurate 
description of the data in the region of low $Q^2$, where 
threshold 
effects appear as $\ln{Q^2/M_c^2}$. In this region heavy quarks are considered 
as generated by the boson-gluon fusion process and the method of~\cite{bgf} 
is used. This method treats the mass logarithms correctly, but fails to 
match the asymptotic limit when 
$Q^2 \gg M_c^2$. The recent work of Lai et. al~\cite{lai}, 
Martin et. al~\cite{mrst} 
and Buza et. al.~\cite{buza} aims to have a smooth transition from the 
threshold region to the asymptotic limit, but such an approach has not yet 
been implemented in the
fitting programs used by the experimental collaborations.
\item {\bf Target mass corrections.} The identification of $x$ with the 
fraction of the nucleon's momentum taken by the struck quark cannot be 
maintained when $Q^2 \simeq M^2$ (where $M$ is the nucleon mass)
and corrections to the formulae of the form $x^2\cdot M^2/Q^2$ 
are necessary~\cite{GPtm}.
\item {\bf Higher Twist effects.} More complicated interactions involving
more than one parton, like secondary interactions of quarks
with the proton remnant, give rise to additional contributions
to the structure functions. These higher twist terms are classified according 
to their $Q^2$ dependence $1/Q^{2n}$, where $n=0$ is leading twist (twist=2), 
$n$=1 is twist=4, etc. They are usually only important at high 
$x$~\cite{virmil}
although higher twist effects at very low $x$ have been considered 
recently~\cite{bartels}.
\item {\bf Nuclear binding effects.} Fixed target DIS experiments
obtain information on the $d$ density by scattering off a deuteron target.
Theoretically the deuteron is often treated as a free neutron and proton; 
binding effects are neglected. Recently much work has been done
in determining more realistic models of the deuteron. For more details 
see Sec.~\ref{sec:bodek}.

\end{itemize}

Estimates of these uncertainties may be obtained by 
repeating several fits varying 
the conditions of each fit in order to ascertain 
the influence of the assumptions.
In the following such a procedure is described with the aim of estimating 
uncertainties in the region of high $x$ and high $Q^2$, relevant to the new 
HERA data.

\subsection{Next-to-Leading Order QCD Fit of NC Data}
The DGLAP evolution equations~\cite{bb.dglap} are solved in the 
next-to-leading order (NLO) $\overline{MS}$ factorisation scheme using the
program QCDNUM~\cite{qcdnum}.  To ensure that the perturbative
approximation is valid a starting scale of \mbox{$Q_0^2=4$ GeV$^2$} is
taken at which four parton densities are parameterized. These are the
up and down valence quarks ($xu_v$ and $xd_v$), the gluon ($xg$), and
the sea quark distribution ($x\rm{S}$).  The sea quark density is
assumed to have equal components from quarks and anti-quarks and is
defined as $x\rm{S}=2x(\bar{u}+\bar{d}+\bar{s}+\bar{c})$ at $Q_0^2$
where $\bar{s}=\bar{u}/2$ and $\bar{d}=\bar{u}$.  Since the kinematic region of
interest is at high $Q^2$, a massless approach for heavy quarks 
is used. The $xb$ density is evolved from zero at a $Q^2$ threshold
defined such that $xb(x,\QQ)=0$ for \QQ$<(5)^2$~${\rm GeV^2}$ and the
charm component is normalised to 2\% of the sea quark density at
$Q^2=Q_0^2$, which gives a good description of the H1
measurements~\cite{fcharm} of \Ftwo$^{c\bar{c}}$, the charm induced
structure function.
 
The functional forms of the parton densities at $Q^2_0$ are as follows:
\begin{eqnarray}
xu_v(x,Q_0^2) &=& A_ux^{B_u}(1-x)^{C_u}(1+D_ux+E_u\sqrt{x})\nonumber\\
xd_v(x,Q_0^2) &=& A_dx^{B_d}(1-x)^{C_d}(1+D_dx+E_d\sqrt{x})\nonumber\\
xg  (x,Q_0^2) &=& A_gx^{B_g}(1-x)^{C_g}(1+D_gx+E_g\sqrt{x})\nonumber\\
x\rm{S}(x,Q_0^2) &=& A_Sx^{B_S}(1-x)^{C_S}(1+D_Sx+E_S\sqrt{x})\nonumber
\end{eqnarray}

The parameters $A_u$, and $A_d$ are determined 
by applying the valence counting rules and $A_g$ is determined in terms of 
the other normalization parameters through the momentum sum rule.
The strong coupling constant, $\alpha_s$, was
fixed at the value $\alpha_s(M_Z^2)=0.118$. 

Measurements of 
\Ftwo~from NMC~\cite{nmc} and BCDMS~\cite{bcdms} on both proton and
deuteron targets were used in the fit.
In addition the \Ftwo~data from H1~\cite{H194} were included for
\mbox{$Q^2 \leq 120$ GeV$^2$}. 
Since there is no data to constrain the low $x$ behaviour of
valence quarks the parameters $B_u$ and $B_d$ were fixed in the fits to
the values given in the MRSR2 parameterization.
The influence of higher twist effects and 
deuteron binding effects was reduced by requiring the data to have 
\mbox{$Q^2>10$ GeV$^2$}, \mbox{$W^2\geq 20$ GeV$^2$} as well as 
\mbox{$Q^2\geq 20$ GeV$^2$} for \mbox{$x\geq 0.5$}. 
The fixed target data were corrected for target mass corrections using the 
Georgi-Politzer approach~\cite{GPtm}, and deuteron binding effects as 
described in~\cite{Frankfurt}. Data with $x > 0.7$ were discarded,
since in this region target mass corrections and 
deuteron binding effects are considered to be too large for reliable 
correction.
The normalizations of all data sets were left free within the quoted
luminosity uncertainty. The fits were performed using the program
MINUIT~\cite{minuit} which minimised the $\chi^2$ defined with
statistical and uncorrelated systematic errors added in quadrature.

Several fits were performed varying in turn the input assumptions on the fit.

\begin{itemize}
\item $\alpha_s$ was varied by $\pm 0.003$ at the scale $M_Z^2$.
\item The influence of remaining higher twist effects was determined by increasing 
the $Q^2$ cut from 10 GeV$^2$ to 15 GeV$^2$.
\item An uncertainty on the charm density was obtained by varying the the charm 
momentum fraction by factors of 2. 
\item The strange component of the sea was varied by $\pm $25\%.
\item No deuteron corrections were applied to the data.
\end{itemize}
  
\begin{figure}[htb]
  \begin{center}
      \epsfig{file=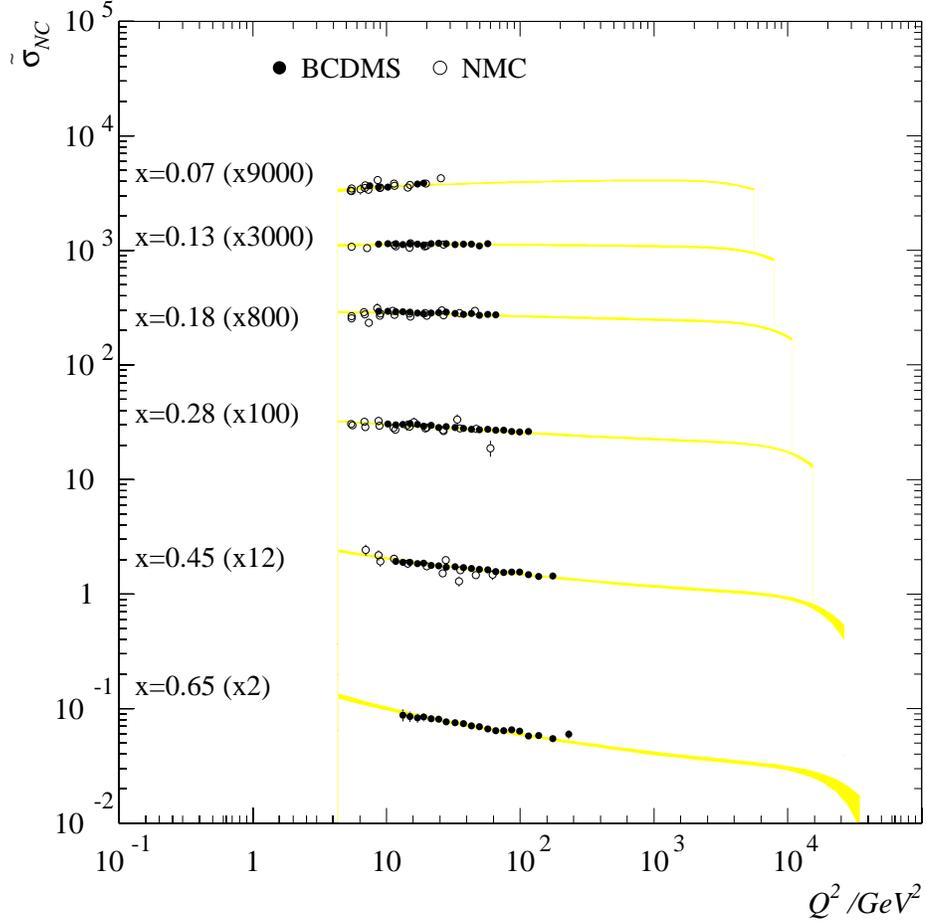,width=0.85\textwidth}
  \end{center}
\caption{$\tilde{\sigma}_{NC}(e^+)$ extrapolated to high $Q^2$. The band
represents the uncertainties arising from variations in the input assumptions
of the fit to NMC, BCDMS and H1 data.}
\label{f2plot}
\end{figure}

The results of the fit are shown in the form of the reduced NC cross section
(this term is defined in Ref.~\cite{mehta}) in Fig~\ref{f2plot}. 
Note that the uncertainty comes purely from the 
variation in the fitted PDFs, all electroweak parameters have been set
to standard values. The total uncertainty on the NC 
cross-section is
at the level of 1\% increasing to 5\% at $x=0.65$. This increase is 
largely due to the
$\alpha_s$ variation and the $Q^2_{min}$ cut applied to the data. 
The variation of the quark sea components has very little effect in this high
$Q^2$ region, however, the effect of the deuteron corrections can be as 
large as 1\%. The present study has considered only `theoretical' 
uncertainties. It is interesting to compare our results with the analysis 
of Botje~\cite{botje}, where experimental 
 systematics are considered (as well as an $\alpha_s$ variation). Botje finds
the uncertainty on $\tilde{\sigma}_{NC}$ to be $4\%$ rising to $11\%$ 
at $x=0.4$ and
$5\%$ to $8\%$ at $x=0.6$. Thus it
seems that at lower $Q^2$ the uncertainties are  dominated by 
experimental errors, whereas at higher $Q^2$ and $x$  theoretical
and experimental uncertainties are of similar size.

\section {Parton  Distributions, $d/u$, and HERA CC Data}
\label{sec:bodek}
\subsection {Extraction of $d/u$ at high $x$ }

\begin{figure}[t]
\centerline{\psfig{figure=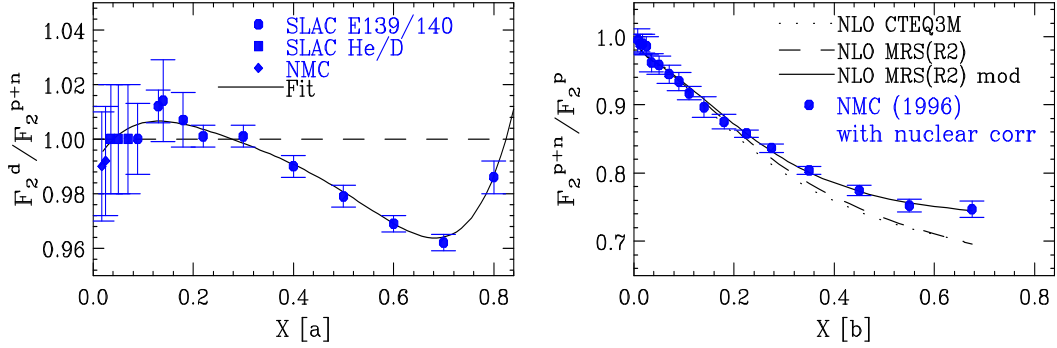,width=5.5in,height=1.8in}}
\caption{[a] The total correction for nuclear 
effects (binding and Fermi motion) in the deuteron,
 $F_2^d/F_2^{n+p}$, as a function of $x$, extracted from fits to
the nuclear dependence of SLAC $F_2$ electron scattering
data. [b] Comparison of NMC $F_2^{n+p}/F_2^p$ (corrected for nuclear
effects) and the prediction in NLO using the  MRS(R2) 
PDF with and without the proposed
modification to the $d/u$ ratio. From ref.~\cite{bodekyang}.}
\label{fig:f2dp}
\end{figure}

The most accurate
information about valence quarks originates from fixed target 
electro/muoproduction data on proton and deuteron targets.
The $u$ valence quark distribution at high $x$
is relatively well
constrained by the proton structure function $F_2^p$.
However, the $d$ valence quark at high $x$ is constrained by the neutron 
structure function $F_2^n$, which is actually extracted from deuteron data.
 Therefore, there is an uncertainty in the $d$ valence quark 
distribution from the corrections for nuclear binding effects in the deuteron. 
In past extractions of $F_2^n$ from deuteron data,
only Fermi motion corrections
were considered, and other 
binding effects were assumed to be negligible.
Recently, corrections for nuclear binding effects in the deuteron,
$F_2^d/F_2^{n+p}$, have been extracted empirically from
fits to the nuclear dependence
of electron scattering data from SLAC experiments E139/140~\cite{GOMEZ}.
The empirical extraction uses a
model proposed by Frankfurt and Strikman~\cite{Frankfurt}
in which all binding effects 
in the deuteron and heavy nuclear targets
are assumed to scale with the nuclear density.
The correction extracted in this empirical way is also in agreement 
(for $x<0.75$) with recent theoretical
calculations
~\cite{duSLAC}
 of nuclear binding effects in the deuteron.
The total correction for nuclear
binding in the deuteron is about 4\% at $x=0.7$
(shown in Fig. \ref{fig:f2dp}[a]), and in a direction which is opposite
to that expected from the previous models which only included
the Fermi motion effects.

\begin{figure}[t]
\centerline{\psfig{figure=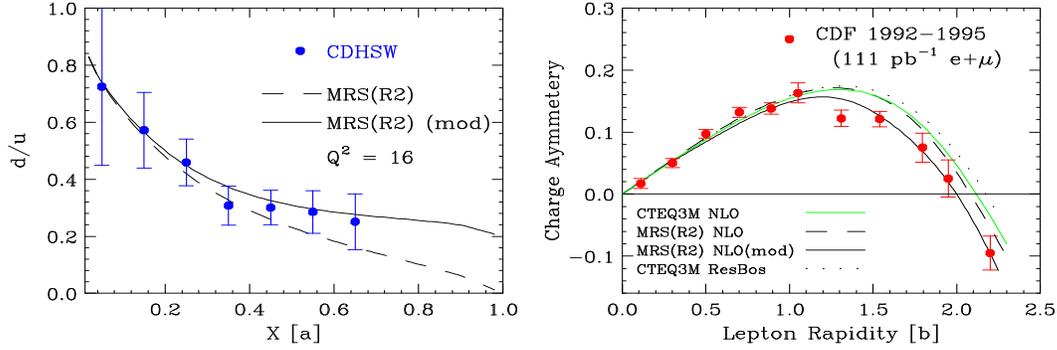,width=5.5in,height=1.8in}}
\caption{[a] The $d/u$ distributions at $Q^2$=$16$ GeV$^2$ 
as a function of $x$ for the standard and modified MRS(R2) PDF
compared to the CDHSW data. 
[b] Comparison of the CDF $W$ asymmetry data with NLO standard
CTEQ3M, MRS(R2), and modified MRS(R2) as a function of the lepton rapidity.
The standard CTEQ3M with 
a resummation calculation is also shown 
for comparison. From ref~\cite{bodekyang}.}
\label{fig:dou}
\end{figure}

The ratio $F_2^d/F_2^p$ is directly related to $d/u$. In leading order QCD,
$2F_2^d/F_2^p -1 \simeq (1+4d/u)/(4+d/u)$ at high $x$.
Bodek and Yang~\cite{bodekyang} have recently performed an NLO analysis on 
the precise NMC $F_2^d/F_2^p$ data~\cite{NMCf2dp} taking deuteron binding
corrections into account.
The ratio $F_2^{p+n}/F_2^p$ is extracted
 by applying the nuclear binding correction
 $F_2^d/F_2^{n+p}$  to the $F_2^d/F_2^p$ data.
 As shown in Fig. \ref{fig:f2dp}[b], the
 standard PDF's
 do not describe this corrected $F_2^{p+n}/F_2^p$.
 Since the $u$ distribution is relatively well constrained,
 Bodek and Yang suggest a correction term to  $d/u$ in the standard PDF's
 (as a function of $x$), which changes the $d$ distribution to fit the data.
 This correction term is  parametrized  
 as a simple quadratic form, $\delta (d/u) = (0.1\pm0.01)(x+1)x$
 for the MRS(R2) PDF,
 where the corrected $d/u$ ratio
 is $(d/u)' = (d/u) + \delta (d/u)$.
 Based on this correction,
 an  MRS(R2)-modified PDF is obtained, as shown in Fig \ref{fig:f2dp}[b].
The correction to other PDF's such as CTEQ3M is similar.
Fig~\ref{fig:dou}[a] illustrates the effect of this modification on the $d/u$
ratio, where we can clearly see that $d/u$ in the
 standard PDF's approaches 0 as $x \rightarrow 1$, whereas 
 the modified $d/u$ ratio  approaches
 $0.2\pm0.02$ as $x \rightarrow 1$, in agreement with a QCD
 prediction~\cite{Farrar}.
Information on $d/u$ which is free from nuclear effects may also be extracted
from  $\nu p$/$\overline{\nu} p$ data, but unfortunately this is rather
inaccurate.  Fig. \ref{fig:dou}[a] shows that  
the CDHSW~\cite{du_cdhsw} data favour the modified PDF's at high $x$.
Further information 
on $d/u$, which is not affected by the corrections
for nuclear effects in the deuteron, can be extracted from $W$ 
production data in hadron colliders.
Fig~\ref{fig:dou}[b] shows that the predicted $W$ asymmetry calculated
with the DYRAD NLO QCD program
using the modified PDF is
in much better agreement with recent CDF data
~\cite{Wasym}
 at large rapidity than standard PDF's.

\begin{figure}
\centerline{\psfig{figure=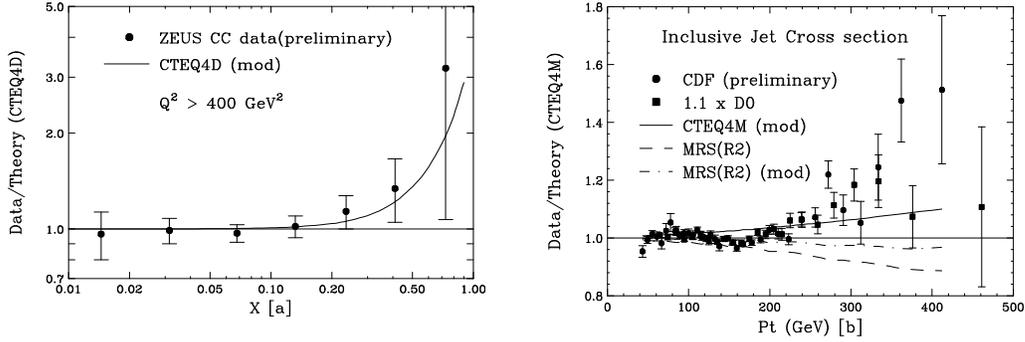,width=5.5in,height=1.8in,clip=1.8in}}

\caption{ [a] The HERA charged current cross section data and 
[b] the CDF and D0 
inclusive jet cross section data are compared 
with both standard and modified PDF's. From ref~\cite{bodekyang}.}
\label{fig:highq2}
\end{figure}

Clearly future $e^+/e^-$ HERA CC data can provide new information 
on the $d/u$ ratio, free from nuclear effects.
When the modified PDF at $Q^2$=$16$ GeV$^2$ is evolved to $Q^2$=$10^4$ GeV$^2$
using the DGLAP NLO equations, it is found that
the modified $d$ distribution at $x=0.5$ is increased by about 40 \% 
in comparison to the standard $d$ distribution.
Fig.~\ref{fig:highq2}[a] shows the 
HERA $e^+$ CC cross section data compared to the predictions of the CTEQ4D
distributions both with and without the modification.
One can easily see that an increased luminosity of such 
data will be crucial in deciding the need for any modification.
Fig.~\ref{fig:highq2}[b] shows that
the modified PDF's also lead to an increase of 10\% in the
QCD prediction for the production rate of very high $P_T$ jets~\cite{jet}
in hadron colliders. 

\subsection {Target mass, higher twist and PDFs at very high $x$}

\begin{figure}[t]
\centerline{\psfig{figure=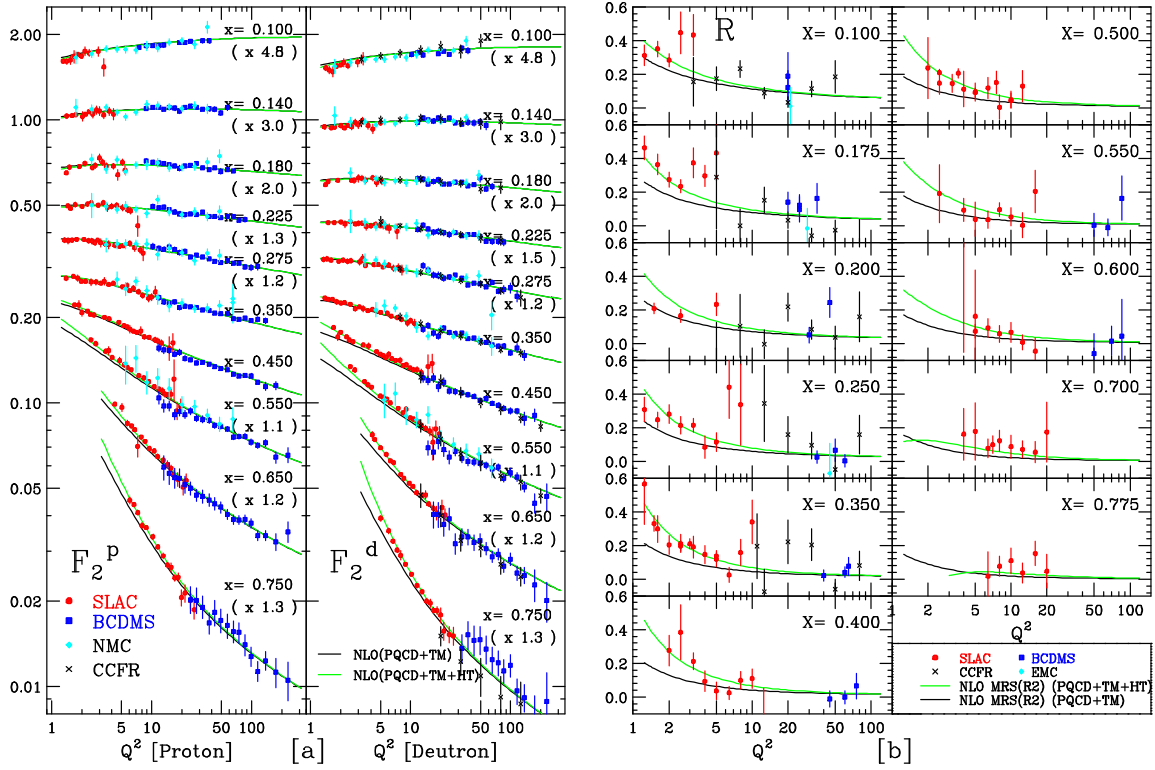,width=6.0in,height=4.0in}}

\caption{Comparison of the modified NLO MRS(R2) PDF plus target mass and
renormalon higher twist corrections, with low $Q^2$ data. 
[a] Comparison of $F_2$ and NLO prediction
with and without higher twist contributions. 
[b] Comparison of $R$ and NLO prediction
with and without 
the higher twist contributions. From ref.~\cite{bodekyang}.}
\label{fig:disht}
\end{figure}

Since all the standard PDF's, including the modified versions, are
fit to data with $x$ less than 0.75, Bodek and Yang also
investigate the validity of the modified MRS(R2) at very high
$x$ ($0.75 < x < 0.98$) by comparing to  $F_2^p$ data at SLAC.
Although the SLAC data at very high $x$ are at reasonable values
of $Q^2$ ($7<Q^2<31$ GeV$^2$), they are in a region in which
 non-perturbative effects such as target mass
 and higher twist are very large.
The Georgi-Politzer calculation is used
~\cite{GPtm}
for the target mass corrections (TM). These involve
using the scaling variable 
$\xi=2x/(1+\sqrt{1+4M^2x^2/Q^2})$ instead of $x$.
Since a complete calculation of higher twist
effects is not available, data at lower $Q^2$ ($1 < Q^2 < 7$ GeV$^2$
and $x < 0.75$) are
used to obtain information on the size of these terms.
Two approaches are used: 
an empirical method and the renormalon model~\cite{renormalon},
 full details of the formalisms are given in~\cite{bodekyang}. 
Both approaches describe the data well. Fig.~\ref{fig:disht} illustrates the
agreement for the renormalon higher twist approach. Thus one has confidence in 
extrapolating these approaches  
to investigate the very high $x$ region.

\begin{figure}[t]
\centerline{\psfig{figure=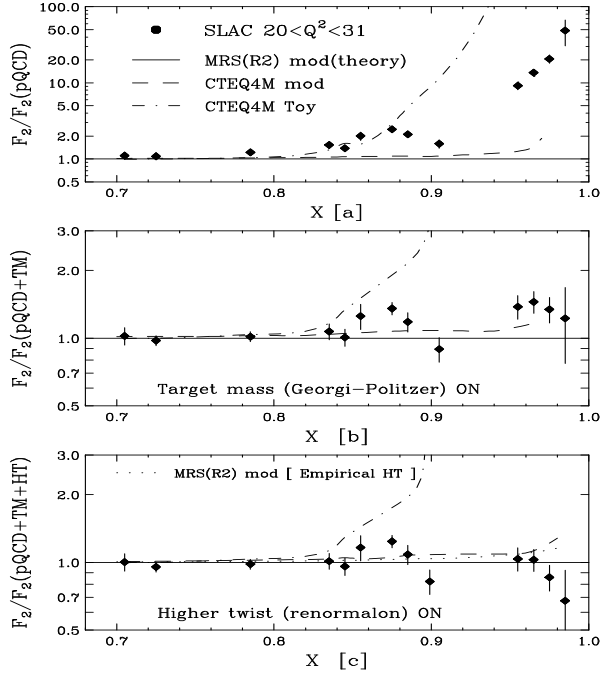,width=3.1in,height=3.5in}}
\caption{Comparison of SLAC $F_2^p$ data
with the predictions of the modified MRS(R2), CTEQ4M and the CTEQ toy model at high $x$
and  $20<Q^2<31$ GeV$^2$. ~~~[a] Ratio to pQCD, [b] ratio to pQCD 
with TM effects,
and [c] ratio to pQCD with TM and higher twist effects. From 
ref.~\cite{bodekyang}.}
\label{fig:highx_highq2}
\end{figure}

There is a wealth of SLAC data
~\cite{SLACres}
 in the region up to $x=0.98$ 
and intermediate $Q^2$ $(7<Q^2<31$ GeV$^2)$. 
Previous PDF fits have not used these data. 
The data for $0.9> x >0.75$ is in the DIS region,
and the data for $x>0.9$ is in the resonance region.
It is worthwhile investigating the resonance region  because
duality arguments
~\cite{Bloom}
indicate that the average behaviour of the resonances and elastic peak
should follow the DIS scaling limit curve.
Fig.~\ref{fig:highx_highq2} shows the ratio of the SLAC data to the predictions
of the modified MRS(R2) at $Q^2$ values ($21<Q^2<30$ GeV$^2$)
where the elastic contribution is negligible.
Fig~\ref{fig:highx_highq2}[c] shows that, with the inclusion of target mass 
and renormalon higher twist
effects, the very high $x$ data from SLAC is remarkably well described 
by the modified MRS(R2) right up to $x=0.98$.  
Fig~\ref{fig:highx_highq2}[c] also illustrates that a good description
is achieved using the empirical estimate of higher twist effects.
Fig.~\ref{fig:highx_highq2} also shows that the CTEQ Toy model
(with an additional 0.5\%
component of $u$ quarks beyond $x>0.75$) overestimates the SLAC data 
by a factor of three at $x = 0.9$ (DIS region).
From these comparisons, it is found that the SLAC $F_2$ data do not support
the CTEQ Toy model
which proposed an additional $u$ quark contribution
at high $x$ as an explanation of the initial HERA high $Q^2$ anomaly
and the CDF high-$P_t$ jet excess.

Thus it seems that there is a need for modification
of current standard PDFs at high $x$, but this conclusion is dependent on the
correctness of deuteron binding calculations. Future 
precise charged current data from HERA will provide
vital information on valence quark densities at high $x$ and high $Q^2$, free
from this uncertainty.

\section{Electroweak Radiative Corrections}
\label{sec:hubert}

Electroweak
radiative corrections of order $O(\alpha)$ have to be taken into
account if a precision of $1\,\%$ is to be achieved in
physics analyses at HERA.
Pure QED corrections contain contributions which are enhanced by
  logarithmic factors like $\log (Q^2/m_e^2)$. In addition, radiation of
  photons can shift kinematic variables from very large to very small
  values inducing additional enhancement factors for NC scattering.
Furthermore, purely weak one-loop corrections will eventually be important for
  a measurement of $M_W$ with a precision of $O(100\,{\rm MeV})$.

As discussed in much detail in the literature \cite{NCcorr}, the
differential cross section for NC scattering when measured in terms of
electron variables receives QED corrections which are large and negative
at small $y_e$ and large $x_e$, whereas at large $y_e$ and small $x_e$
large positive corrections may reach the level of $100\,\%$. These
corrections can be reduced, sometimes considerably, with measurements
based on hadronic or mixed kinematic variables. Self energy corrections
(the running of $\alpha(Q^2)$) are of order 10\,\% and,
finally, purely weak corrections are negligible at low $Q^2$, but
increase to the level of a few per cent at very large $Q^2$. 

In the following we give a more detailed discussion of electroweak
radiative corrections to the charged current process
\cite{CCcorr1,CCcorr2,CCcorr3}, firstly since their treatment is more
involved and has been less well described in the literature than the NC
case; secondly because the implementations in available numerical
programs appear to have larger uncertainties than the corresponding NC
codes.

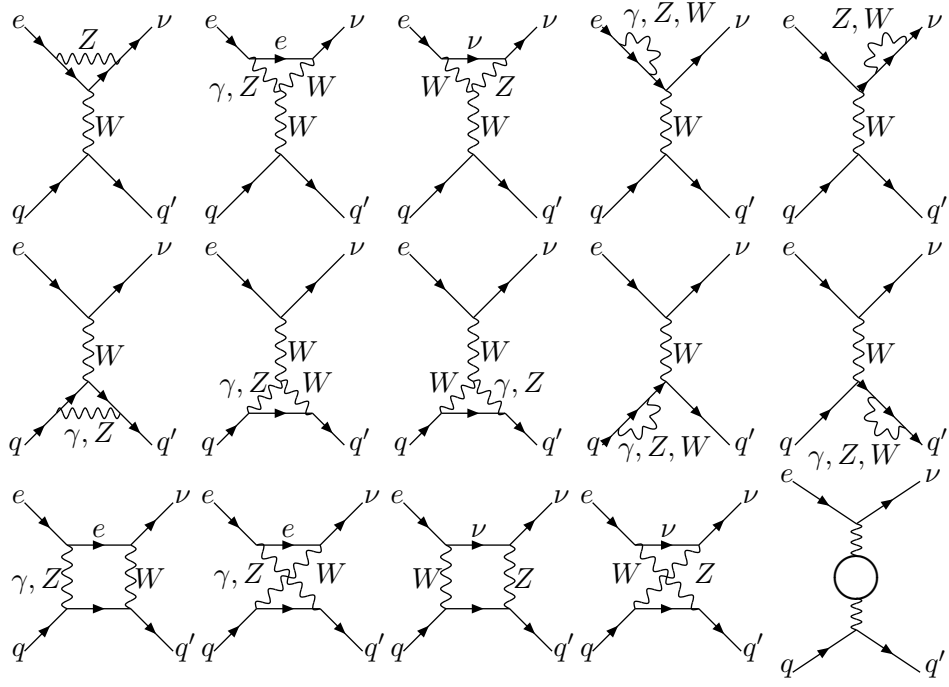
\begin{figure}[htbp]
\small
%
\begin{picture}(75,80)(-75,0)
\ArrowLine(8,76)(20,64)
\ArrowLine(20,64)(32,52)
\ArrowLine(32,52)(44,64)
\ArrowLine(44,64)(56,76)
\Photon(32,52)(32,28){2}{4}
\ArrowLine(8,4)(32,28)
\ArrowLine(32,28)(56,4)
\Photon(20,64)(44,64){2}{5}
\put(3.2,76){$e$}
\put(58,76){$\nu$}
\put(3.2,4){$q$}
\put(58,4){$q'$}
\put(34.4,35.2){$W$}
\put(28,68){$Z$}
\end{picture}
%
\begin{picture}(75,80)(-55,0)
\ArrowLine(8,76)(20,64)
\Photon(20,64)(32,52){2}{3}
\Photon(32,52)(44,64){2}{3}
\ArrowLine(44,64)(56,76)
\Photon(32,52)(32,28){2}{4}
\ArrowLine(8,4)(32,28)
\ArrowLine(32,28)(56,4)
\ArrowLine(20,64)(44,64)
\put(3.2,76){$e$}
\put(58,76){$\nu$}
\put(3.2,4){$q$}
\put(58,4){$q'$}
\put(5,50){$\gamma,Z$}
\put(41.6,50){$W$}
\put(34.4,35.2){$W$}
\put(29.6,68){$e$}
\end{picture}
%
\begin{picture}(75,80)(-35,0)
\ArrowLine(8,76)(20,64)
\Photon(20,64)(32,52){2}{3}
\Photon(32,52)(44,64){2}{3}
\ArrowLine(44,64)(56,76)
\Photon(32,52)(32,28){2}{4}
\ArrowLine(8,4)(32,28)
\ArrowLine(32,28)(56,4)
\ArrowLine(20,64)(44,64)
\put(3.2,76){$e$}
\put(58,76){$\nu$}
\put(3.2,4){$q$}
\put(58,4){$q'$}
\put(12,50){$W$}
\put(40,50){$Z$}
\put(34.4,35.2){$W$}
\put(29.6,68){$\nu$}
\end{picture}
%
\begin{picture}(75,80)(-15,0)
\ArrowLine(8,76)(12,72)
\ArrowLine(12,72)(28,56)
\ArrowLine(28,56)(32,52)
\PhotonArc(20,64)(8,-45,135){2}{4}
\ArrowLine(32,52)(56,76)
\Photon(32,52)(32,28){2}{4}
\ArrowLine(8,4)(32,28)
\ArrowLine(32,28)(56,4)
\put(3.2,76){$e$}
\put(58,76){$\nu$}
\put(3.2,4){$q$}
\put(58,4){$q'$}
\put(34.4,35.2){$W$}
\put(16,78){$\gamma,Z,W$}
\end{picture}
%
\begin{picture}(75,80)(5,0)
\ArrowLine(8,76)(32,52)
\ArrowLine(32,52)(36,56)
\ArrowLine(36,56)(52,72)
\ArrowLine(52,72)(56,76)
\PhotonArc(44,64)(8,45,225){2}{4}
\Photon(32,52)(32,28){2}{4}
\ArrowLine(8,4)(32,28)
\ArrowLine(32,28)(56,4)
\put(3.2,76){$e$}
\put(58,76){$\nu$}
\put(3.2,4){$q$}
\put(58,4){$q'$}
\put(34.4,35.2){$W$}
\put(21.6,76){$Z,W$}
\end{picture}
%
%
\begin{picture}(75,80)(-75,5)
\ArrowLine(8,76)(32,52)
\ArrowLine(32,52)(56,76)
\Photon(32,52)(32,28){2}{4}
\ArrowLine(8,4)(20,16)
\ArrowLine(20,16)(32,28)
\ArrowLine(32,28)(44,16)
\ArrowLine(44,16)(56,4)
\Photon(20,16)(44,16){2}{5}
\put(3.2,76){$e$}
\put(58,76){$\nu$}
\put(2,2){$q$}
\put(58,2){$q'$}
\put(34.4,34.4){$W$}
\put(23,5){$\gamma,Z$}
\end{picture}
%
\begin{picture}(75,80)(-55,5)
\ArrowLine(8,76)(32,52)
\ArrowLine(32,52)(56,76)
\Photon(32,52)(32,28){2}{4}
\ArrowLine(8,4)(20,16)
\ArrowLine(20,16)(44,16)
\ArrowLine(44,16)(56,4)
\Photon(20,16)(32,28){2}{3}
\Photon(32,28)(44,16){2}{3}
\put(3.2,76){$e$}
\put(58,76){$\nu$}
\put(2,2){$q$}
\put(58,2){$q'$}
\put(8.8,24){$\gamma,Z$}
\put(40,24){$W$}
\put(34.4,36){$W$}
\end{picture}
%
\begin{picture}(75,80)(-35,5)
\ArrowLine(8,76)(32,52)
\ArrowLine(32,52)(56,76)
\Photon(32,52)(32,28){2}{4}
\ArrowLine(8,4)(20,16)
\ArrowLine(20,16)(44,16)
\ArrowLine(44,16)(56,4)
\Photon(20,16)(32,28){2}{3}
\Photon(32,28)(44,16){2}{3}
\put(3.2,76){$e$}
\put(58,76){$\nu$}
\put(2,2){$q$}
\put(58,2){$q'$}
\put(40,24){$\gamma,Z$}
\put(15.2,24){$W$}
\put(34.4,36){$W$}
\end{picture}
%
\begin{picture}(75,80)(-15,5)
\ArrowLine(8,76)(32,52)
\ArrowLine(32,52)(56,76)
\Photon(32,52)(32,28){2}{4}
\ArrowLine(8,4)(12,8)
\ArrowLine(12,8)(28,24)
\ArrowLine(28,24)(32,28)
\PhotonArc(20,16)(8,225,45){2}{4}
\ArrowLine(32,28)(56,4)
\put(3.2,76){$e$}
\put(58,76){$\nu$}
\put(2,2){$q$}
\put(58,2){$q'$}
\put(34.4,35.2){$W$}
\put(14,0){$\gamma,Z,W$}
\end{picture}
%
\begin{picture}(75,80)(5,5)
\ArrowLine(8,76)(32,52)
\ArrowLine(32,52)(56,76)
\Photon(32,52)(32,28){2}{4}
\ArrowLine(8,4)(32,28)
\ArrowLine(32,28)(36,24)
\ArrowLine(36,24)(52,8)
\ArrowLine(52,8)(56,4)
\PhotonArc(44,16)(8,135,315){2}{4}
\put(3.2,76){$e$}
\put(58,76){$\nu$}
\put(2,2){$q$}
\put(58,2){$q'$}
\put(34.4,35.2){$W$}
\put(13,-3){$\gamma,Z,W$}
\end{picture}
%
%
\begin{picture}(75,64)(-79,6)
\ArrowLine(4,64)(20,48)
\ArrowLine(20,48)(44,48)
\ArrowLine(44,48)(60,64)
\Photon(20,48)(20,24){2}{4}
\Photon(44,48)(44,24){2}{4}
\ArrowLine(4,8)(20,24)
\ArrowLine(20,24)(44,24)
\ArrowLine(44,24)(60,8)
\put(-0.8,64){$e$}
\put(29.6,51.2){$e$}
\put(60.8,64){$\nu$}
\put(-0.8,6.4){$q$}
\put(60.8,6.4){$q'$}
\put(-1,32){$\gamma,Z$}
\put(45.6,32){$W$}
\end{picture}
%
\begin{picture}(75,64)(-72,6)
\ArrowLine(4,64)(20,48)
\ArrowLine(20,48)(44,48)
\ArrowLine(44,48)(60,64)
\Photon(20,48)(44,24){2}{5}
\Photon(44,48)(20,24){2}{5}
\ArrowLine(4,8)(20,24)
\ArrowLine(20,24)(44,24)
\ArrowLine(44,24)(60,8)
\put(-0.8,64){$e$}
\put(29.6,51.2){$e$}
\put(60.8,64){$\nu$}
\put(-0.8,6.4){$q$}
\put(60.8,6.4){$q'$}
\put(4,34){$\gamma,Z$}
\put(42.4,34){$W$}
\end{picture}
%
\begin{picture}(75,64)(-65,6)
\ArrowLine(4,64)(20,48)
\ArrowLine(20,48)(44,48)
\ArrowLine(44,48)(60,64)
\Photon(20,48)(20,24){2}{4}
\Photon(44,48)(44,24){2}{4}
\ArrowLine(4,8)(20,24)
\ArrowLine(20,24)(44,24)
\ArrowLine(44,24)(60,8)
\put(-0.8,64){$e$}
\put(29.6,51.2){$\nu$}
\put(60.8,64){$\nu$}
\put(-0.8,6.4){$q$}
\put(60.8,6.4){$q'$}
\put(7.2,32){$W$}
\put(45.6,32){$Z$}
\end{picture}
%
\begin{picture}(75,64)(-58,6)
\ArrowLine(4,64)(20,48)
\ArrowLine(20,48)(44,48)
\ArrowLine(44,48)(60,64)
\Photon(20,48)(44,24){2}{5}
\Photon(44,48)(20,24){2}{5}
\ArrowLine(4,8)(20,24)
\ArrowLine(20,24)(44,24)
\ArrowLine(44,24)(60,8)
\put(-0.8,64){$e$}
\put(29.6,51.2){$\nu$}
\put(60.8,64){$\nu$}
\put(-0.8,6.4){$q$}
\put(60.8,6.4){$q'$}
\put(11,34){$W$}
\put(42.4,34){$Z$}
\end{picture}
%
\begin{picture}(75,80)(-51,10)
\ArrowLine(8,76)(32,60)
\ArrowLine(32,60)(56,76)
\Photon(32,60)(32,48){2}{3}
\Photon(32,32)(32,20){2}{3}
\ArrowLine(8,4)(32,20)
\ArrowLine(32,20)(56,4)
\thicklines\put(32,40){\circle{16}}
\put(3.2,76){$e$}
\put(58,76){$\nu$}
\put(3.2,4){$q$}
\put(58,4){$q'$}
\end{picture}
\caption{One-loop Feynman diagrams for $eq \rightarrow \nu q'$
  scattering.} 
\label{rcfig1}
\end{figure}

\begin{figure}[htbp]
%
\begin{picture}(80,90)(-75,-5)
\ArrowLine(10,75)(25,65)
\ArrowLine(25,65)(40,55)
\Photon(25,65)(50,82){2}{4}
\ArrowLine(40,55)(70,75)
\Photon(40,55)(40,25){2}{4}
\ArrowLine(10,5)(40,25)
\ArrowLine(40,25)(70,5)
\put(1,75){$l_{\mu}$}
\put(73,75){$l'_{\mu}$}
\put(-1,5){$q_{\mu}$}
\put(73,5){$q'_{\mu}$}
\put(50,84){$k_{\mu}$}
\put(32,-10){(a)}
\end{picture}
\begin{picture}(80,90)(-85,-5)
\ArrowLine(10,75)(40,55)
\ArrowLine(40,55)(70,75)
\Photon(40,55)(40,25){2}{4}
\ArrowLine(10,5)(25,15)
\ArrowLine(25,15)(40,25)
\Photon(25,15)(50,0){2}{4}
\ArrowLine(40,25)(70,5)
\put(32,-10){(b)}
\end{picture}
\begin{picture}(80,90)(-95,-5)
\ArrowLine(10,75)(40,55)
\ArrowLine(40,55)(70,75)
\Photon(40,55)(40,25){2}{4}
\ArrowLine(10,5)(40,25)
\ArrowLine(40,25)(55,15)
\ArrowLine(55,15)(70,5)
\Photon(55,15)(80,30){2}{4}
\put(32,-10){(c)}
\end{picture}
\begin{picture}(80,90)(-105,-5)
\ArrowLine(10,75)(40,55)
\ArrowLine(40,55)(70,75)
\Photon(40,55)(40,25){2}{4}
\Photon(40,40)(75,40){2}{4}
\ArrowLine(10,5)(40,25)
\ArrowLine(40,25)(70,5)
\put(32,-10){(d)}
\end{picture}
\caption{Feynman diagrams for radiative charged current scattering $eq
  \rightarrow \nu q' \gamma$ and notation of 4-momenta for external
  particles.}  
\label{rcfig2}
\end{figure}
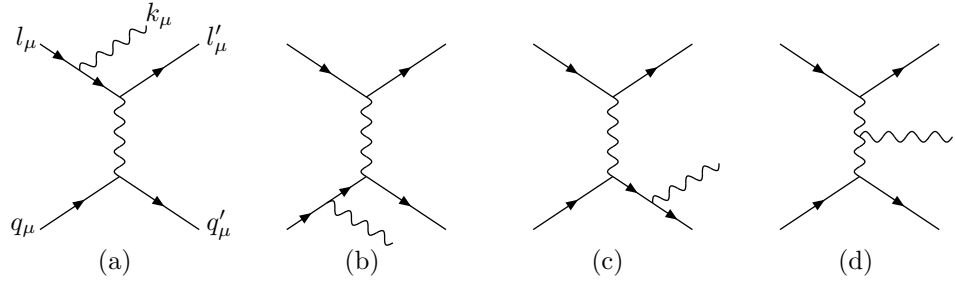

Fig.~\ref{rcfig1} shows the complete set of electroweak one-loop Feynman 
diagrams 
contributing to $eq_f \rightarrow \nu q_{f'}$ scattering. The list
comprises corrections to the $e\nu W$ and $q_f q_{f'} W$ vertices, self
energy corrections to the external lines, graphs for the $W$ self energy 
and box diagrams. Diagrams with an additional photon in the loop gives
rise to infrared divergent contributions. They can be regularized, for
example by using a finite mass for the photon, and cancel against
similarly infrared divergent contributions from radiative CC scattering
$eq_f \rightarrow \nu q_{f'}\gamma$ (see Fig.~\ref{rcfig2}) when calculating 
the cross section.

The calculation is more complicated than the one for NC scattering since
for the CC case there is no simple enough subset of Feynman diagrams
which is gauge invariant. In particular, all four diagrams of Fig.~\ref{rcfig2}
describing the emission of a photon from the incoming lepton (a), from
the incoming (b) and outgoing quark (c) and from the internal $W$ (d)
are necessary to obtain a physically meaningful result.

Internal fermion propagators in the diagrams of Fig.~\ref{rcfig2} give rise to
terms with pole factors which become large when the photon is collinear
with one of the external charged fermions. The differential cross
section for radiative CC scattering can thus be written as
\begin{equation}
\left.\frac{d^5\sigma}{dxdQ^2d^3k}\right|_{eq\rightarrow\nu q'\gamma} =
\frac{A_l}{kl} + \frac{A_q}{kq} + \frac{A_{q'}}{kq'} + {\rm
  non-pole~terms}, 
\label{rc1}
\end{equation}
where the notation shown in Fig.~\ref{rcfig2}(a) for the particle's 4-momenta 
was used. This shows that the numerical calculation of the
$\gamma$-inclusive cross section has to cope with nearly-divergent
behavior in different phase space regions. Note that the separation
into pole factors in Eq.~\ref{rc1} is not unique: finite (i.e.\ non-pole)
terms can be shifted freely between the three contributions. A specific
separation of the radiative cross section like that given in Eq.~\ref{rc1} 
is needed for
the construction of a well-behaved numerical algorithm, but the
individual contributions do not have a well-defined physical meaning
when taken separately.

To obtain a gauge invariant, i.e.\ physically meaningful separation, one
can organize the terms contributing to the complete cross section
according to their dependence on the electric charge of the incoming
particles. Replacing the charge of the $W^{\pm}$ by that of the incoming
lepton, $e_l = e_W$, and that of the final quark $q_{f'}$ by that of the
incoming quark $q_f$ and the lepton, $e_{f'} = e_f + e_l$, one can write
\begin{equation}
\left.\frac{d^5\sigma}{dxdQ^2d^3k}\right|_{eq\rightarrow\nu q'\gamma} =
e_l^2 I_{\rm lep} + e_l e_f I_{\rm int} + e_f^2 I_{\rm qua}.
\label{rc2}
\end{equation}
The individual parts in this separation are gauge invariant and due to
their association to fermion charges it is justified to call them
``leptonic'', ``interference'', and ``quarkonic'', as indicated by the
subscripts in Eq.~\ref{rc2}. A closer inspection of the diagrams in 
Fig.~\ref{rcfig2} 
shows that $I_{\rm lep}$ contains pole terms $\propto 1/kl$ and
$1/kq'$, whereas $I_{\rm qua}$ contains terms $\propto 1/kq$ and
$1/kq'$. The interference contribution $I_{\rm int}$ receives
contributions from all three pole factors $1/kl$, $1/kq$ and $1/kq'$.
After integration over the photon momentum, the pole factors give rise
to logarithms of the corresponding fermion masses: $\log (Q^2/m_e^2)$,
$\log (Q^2/m_{q_f}^2)$ and $\log (Q^2/m_{q_{f'}}^2)$. Here we assume
that collinear singularities are regularized by introducing finite
masses for quarks.

The virtual corrections from Fig.~\ref{rcfig1} 
can be separated in the same way.
After combining real and virtual corrections, not only do infrared
divergent contributions cancel (allowing the removal of the
regularization by the finite photon mass), but also many mass logarithms
disappear in the complete final result: writing the complete 
$O(\alpha)$-corrected cross section similarly to Eq.~\ref{rc2}:
\begin{equation}
\left.\frac{d^5\sigma}{dxdQ^2d^3k}\right|_{eq\rightarrow\nu q'(+\gamma)} =
e_l^2 J_{\rm lep} + e_l e_f J_{\rm int} + e_f^2 J_{\rm qua},
\label{rc3}
\end{equation}
one finds that $i$) all logarithms related to the final quark disappear, 
i.e.\ $J_{\rm lep}$ contains $\log (Q^2/m_e^2)$ terms only, 
$J_{\rm qua}$ contains $\log (Q^2/m_{q_f}^2)$ terms only 
and $ii$) all mass logarithms
cancel in $J_{\rm int}$, i.e.\ the interference contribution contains no 
mass singularities. Therefore the effect of photonic corrections on the
cross section for CC scattering can be summarized as follows:
\begin{itemize}
\item Leptonic contributions proportional to $e_l^2$ contain terms which 
  are enhanced by the large logarithm $\log (Q^2/m_e^2)$. This
  contribution has a dependence on kinematic variables which is very
  similar to the NC case except that the enhancement at large $y_e$ is
  much less pronounced because of the much weaker dependence of the CC
  Born cross section on $Q^2$. 
\item Interference contributions are of order $O(\alpha/\pi)$ without
  logarithmic enhancement. 
\item Quarkonic contributions contain uncancelled $\log (Q^2/m_{q_f}^2)$
  terms. These mass singularities have the same structure as those
  appearing in a calculation of $O(\alpha_{\rm s})$ QCD corrections due
  to the emission of gluons from incoming and scattered quarks (in
  dimensional regularization one would find the same $1/\epsilon$
  divergences apart from differing charge factors: $\alpha
  e_f^2$ instead of $\alpha_{\rm s} C_F$). As with QCD corrections, the
  mass singularities due to QED corrections can be factorized and
  absorbed into renormalized parton distribution functions. Thereby the
  final results do not depend any more on a regularization parameter
  (like the quark mass), but the $Q^2$-evolution equations for parton
  distribution functions receive an additional QED term:
\begin{eqnarray}
\displaystyle \fl
Q^2\frac{d}{dQ^2} q_f(x,Q^2) = 
\frac{\alpha_{\rm s}(Q^2)}{2\pi} \int_0^1 \frac{dz}{z} 
\left(C_F P_{f/f}(z) q_f\left(\frac{x}{z},Q^2\right) 
+ P_{f/g}(z) g\left(\frac{x}{z},Q^2\right) \right) \nonumber \\
\displaystyle
+ \frac{\alpha(Q^2)}{2\pi} e_f^2 \int_0^1 \frac{dz}{z} 
P_{f/f}(z) q_f\left(\frac{x}{z},Q^2\right). 
\label{rc4}
\end{eqnarray}
Numerical estimates \cite{RCquark} show that such effects reach the level of 
$1\,\%$ only at very large $x$ and $Q^2$. 
\end{itemize}
One should note that the available numerical programs either simply
ignore the quarkonic contributions or suppress them by using a
relatively large value for the quark mass (of the order of several GeV).
Another level of approximation justified by the above discussion is to
neglect the interference contribution. Numerical results for the
individual contributions have been given in \cite{CCcorr1,CCcorr3}.

Having separated the purely photonic corrections as described above,
there still remain purely weak contributions (most prominently, but not
exclusively, due to the $W$ self energy) which can be combined into
over-all form factors~\cite{CCcorr1,EWhera}. For electron scattering with
unpolarized beams Eq.~\ref{eq:empCC} becomes:
\begin{eqnarray}
\displaystyle \fl
\left.\frac{d\sigma}{dxdQ^2}\right|^{CC}_{e^-p} = 
\frac{G_{\mu}^2}{2\pi} \left(\frac{M_W^2}{Q^2 + M_W^2}\right)^2 
\Bigg\{ \Bigl(\rho_{CC}^{eu}(x,Q^2)\Bigr)^2 \sum_{f = u,c} q_f(x,Q^2) 
\nonumber \\ \displaystyle 
+ (1-y)^2\left(\rho_{CC}^{ed}(x,Q^2)\right)^2 \sum_{f = d,s}
\bar{q}_f(x,Q^2) \Bigg\}. 
\label{rc5}
\end{eqnarray}
The form factors $\rho_{CC}^{ef}(x,Q^2)$ depend on $Q^2$ and
$x$ as well as on the type of the scattered quark (due to the presence
of box diagrams). The above expression assumes that the CC cross section
is normalized with the help of the $\mu$-decay constant $G_{\mu}$.

The cross section also depends explicitly on the $W$ boson mass and,
through the form factors $\rho_{CC}^{ef}$, on all other parameters of
the electroweak standard model.

\subsection{Choice of independent parameters}

One possible set of independent parameters is $\alpha$, 
$m_t$ (the top mass), $M_H$ (the Higgs mass), $M_W$,
$M_Z$ (in addition, there are light
fermion masses and CKM matrix elements, which we do not consider). 
A special role is played by
$\alpha$ which is kept fixed since it is well-measured and is a
parameter of QED which is embedded in the full electroweak theory. 
The Higgs mass is also special: because of the weak dependence on
$M_H$ in present day experiments, one can keep it as an
external parameter and investigate the effect of changing it separately
(see Sec.~\ref{sec:ken}). Thus the
essential parameters are $M_W$, $M_Z$ and $m_t$, since $G_{\mu}$
is related to these parameters by the SM expression given in Eq.~\ref{rc6}.

We consider the following two possible scenarios:
\begin{itemize}
\item Choose $G_{\mu}$, $M_W$ and $M_Z$ as independent parameters. Then
  $m_t$ is a prediction and a specific choice of values for
  $G_{\mu}$, $M_W$ and $M_Z$ will predict a value for $m_t$ (possibly in
  disagreement with experimental results). 
  Fixing $G_{\mu}$ and $M_Z$ to their measured values (from $\beta$ decay and
  LEP experiments, respectively) the
  CC cross section depends on $M_W$ and a measurement at HERA can be
  interpreted as a measurement of the $W$ mass. Note that this assumes
  the validity of the Standard Model and the measurement is therefore
  not a direct $W$-mass measurement. (This is also true for scenario 2).
  Alternatively, keeping $M_W$ and
  $M_Z$ fixed, the charged current cross section could be interpreted as
  a measurement of $G_{\mu}$. This is interesting since it allows a
  direct comparison of the high-$Q^2$ experiments at HERA with the
  low-energy measurement of $\mu$ decay
                  
\item Choose $M_W$, $M_Z$ and $m_t$ as independent parameters (on-shell
  scheme). Then $G_{\mu}$ is a prediction of the theory given by
\begin{equation}
G_{\mu} = \frac{\pi\alpha}{\sqrt{2}s_{\rm w}^2 M_W^2} 
\frac{1}{1-\Delta r}
~~~ {\rm with} ~~~ s_{\rm w}^2 = 1 - \frac{M_W^2}{M_Z^2}
\label{rc6}
\end{equation}
and $\Delta r$, which embodies the radiative corrections to the $\mu$-decay, 
is a function of $M_W$, $M_Z$, $m_t$ (and $\alpha$, $M_H$, 
of course). $\Delta r$ has to be taken into account in a consistent
$O(\alpha)$ prescription and should be combined with the other
$O(\alpha)$ corrections in the CC cross section formula, i.e.\ replacing 
\begin{equation}
\rho_{CC}^{ef} \longrightarrow \frac{\rho_{CC}^{ef}}{(1 - \Delta
  r)^2} = \hat{\rho}_{CC}^{ef}.
\label{rc7}
\end{equation}
Because of the inclusion of $\Delta r$ the
new form factors have a strong dependence on $m_t$ since
\begin{equation}
\Delta r = \Delta \alpha - \frac{c_w^2}{s_{\rm w}^2}\Delta \bar{\rho} 
+ \Delta r^{\rm rem} ~~~ {\rm with} ~~~ \Delta \bar{\rho} 
= \frac{3G_{\mu} m_t^2}{8\pi^2\sqrt{2}}. 
\label{rc8}
\end{equation}
Furthermore the dependence of the CC cross section on $M_W$ is much stronger
due to the replacement
\begin{equation}
\left(\frac{m_W^2}{Q^2+m_W^2}\right)^2 \longrightarrow
\left(\frac{1}{1-m_W^2/m_Z^2} \frac{1}{Q^2 + m_W^2}\right)^2. 
\label{rc9}
\end{equation}
Applied to the situation at HERA, keeping $M_Z$ and $m_t$ at
their experimental values, the measurement of the CC cross section can once
more be interpreted as an indirect measurement of 
$M_W$~\cite{EWheraB,Ref:Elsen} which is, through
Eq.~\ref{rc6}, equivalent to a measurement of $G_{\mu}$, however in a
process at large momentum transfer.
\end{itemize}
Both schemes are theoretically equivalent, a specific choice should be
motivated by the aim of the analysis.  We repeat: the
above discussion focussed on the essential parameters $M_W$, $M_Z$,
$m_t$, but, in addition, a weak dependence on $M_H$ is always
present, both in $\Delta r$ as well as in the form factors
$\rho_{CC}^{ef}$, this is quantified further in Sec.~\ref{sec:ken}

It might be instructive to see how the situation simplifies in the case
 when radiative corrections are neglected altogether. Then $m_t$ and
 $M_H$ do not play any role for CC DIS. In scenario 1 where
 $G_{\mu}$ and $M_W$ are considered as independent parameters, $M_Z$
 is also irrelevant. With fixed $G_{\mu}$, the measured cross section can be
 fitted by varying $M_W$. The resulting value has been called a `propagator
 mass' in previous experimental literature since it is due to the
 dependence of the cross section on $M_W$ via the propagator
 only. Keeping $M_W$ fixed leaves no possibility to adjust the $Q^2$
 dependence, and then the measurement only determines the
 normalization (in terms of $G_{\mu}$) of the cross section. 
 Scenario 2, with fixed $M_Z$, allows us to combine the sensitivity to the
 propagator and to the normalization in one common parameter, i.e.\ the
 $W$ mass, as made explicit in Eq.~\ref{rc9}. In Sec.~\ref{sec:ken} 
we consider  both scenarios.

\subsection{Numerical results and discussion of uncertainties}

\begin{figure}[hptb] 
\unitlength 1mm
\begin{picture}(120,90)
\put(31,-6){\epsfxsize=12cm \epsfysize=10cm
\epsfbox{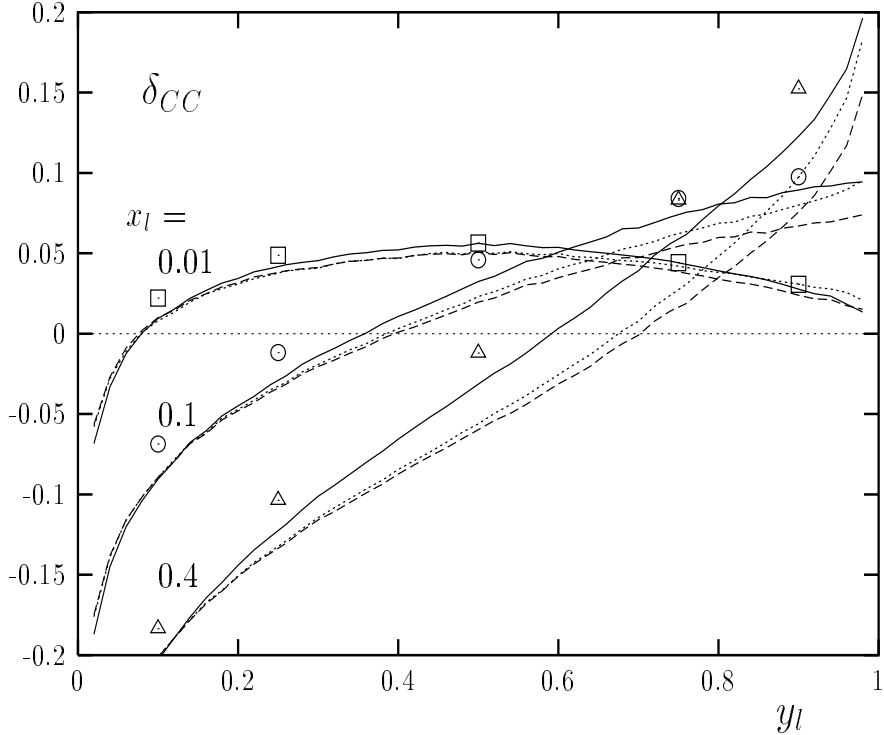}}
\end{picture}
\caption{\it Comparison of radiative corrections for the charged current
  cross section, $\delta_{CC} = \left.d\sigma/dx_ldy_l
  \right|_{O(\alpha)} / \left.d\sigma/dx_ldy_l \right|_{\rm Born} - 1$,
  as a function of $y_l$ for fixed values of $x_l$. The full curves
  include the complete QED and weak corrections, the dashed curves are
  without purely weak corrections and the dotted lines show the leptonic
  QED corrections only (all from {\tt epcctot}). Open symbols are
  obtained from {\tt DJANGOH} and include leptonic QED and purely weak
  corrections.}
\label{rcfig3}
\end{figure}

The presently available numerical programs for the calculation of the CC
cross section do not all take into account the complete set of one-loop
and one-photon corrections described above. Two programs with the
complete set of $O(\alpha)$ corrections ({\tt DISEPW} \cite{CCcorr1} and
{\tt epcctot} \cite{hseprc}) have been compared in the HERA workshop
\cite{NCcorr} and found to agree well except in the phase space corners
of small $x$, small $y$ or large $x$, all $y$. These programs are,
however, not very well-suited for an application to realistic
experiments since they do not allow to apply experimental cuts and,
secondly, since they are restricted to the use of leptonic variables
whereas experiments have to use hadronic variables.
The Monte Carlo event generator {\tt HERACLES/DJANGOH}~\cite{djangoh} 
circumvents these
two restrictions, however, it has the CC radiative corrections
implemented in an approximation where photonic radiative corrections
from lepton-quark interference and from quark radiation are neglected.
Theoretically this is motivated to the extent that an
accuracy of $1\%$ is sufficient, as discussed above. 

Fig.~\ref{rcfig3} shows a comparison of
results obtained with the most recent version of {\tt DJANGOH} with the
complete $O(\alpha)$ results from {\tt epcctot} for lepton variables. At
small values of $x_e$, the comparison is satisfactory, but at larger
values of $x_e$, the Monte Carlo results lie above those from {\tt
  epcctot}. Here improvements are still needed. More detailed comparisons, 
in particular between {\tt DJANGOH}
and {\tt HECTOR} \cite{hector}, are presently being performed \cite{heramc}.

Apart from the approximations in available program implementations,
there are additional intrinsic uncertainties of predictions for the CC
cross section. First, a theoretically consistent treatment of quarkonic
radiation as described above is not available yet. The uncertainty from
this source is most likely below $1\%$, 
and is thus negligible except at very large $Q^2$~\cite{RCquark}.
Secondly, higher-order corrections are to be expected, first of all from
multi-photon effects, but also from combined QED--QCD corrections of
$O(\alpha \alpha_{\rm s})$ and possibly from two-loop weak
corrections. Multi-photon effects due to soft photons can be treated by
exponentiation. This effect can be taken into account only in {\tt epcctot} or
{\tt HECTOR}. It is expected to be numerically significant at
low $y_e$, low $x_e$ ($O(2\%)$ for $x_e<0.1$) and very large $x_e$ ($O(10\%)$ 
for $x_e>0.9$)~\cite{CCcorr3}, i.e. close to the phase space boundaries. 
Two-photon contributions of $O(\alpha^2)$ in the
leading logarithmic approximation are also known and can be calculated
with the help of {\tt HECTOR} but no numerical evaluation of this contribution
has been performed for CC scattering. $O(\alpha \alpha_{\rm s})$
corrections are not expected to become important. An indication of this comes
from the fact that in NC scattering the larger $O(\alpha_{\rm s})$ correction 
to $F_3$ stays below 1.5\%, for values of $x$ and $Q^2$ relevant at HERA.
Two-loop weak contributions are also not 
expected to be important for the cross section itself.  However,
two-loop effects can affect the value of $M_W$ at the level of several
tens of MeV when determined from $G_{\mu}$ via Eq.~\ref{rc6}. The
dominating higher-than one-loop effects to the $G_{\mu} - M_W$ relation
are implemented in all availabe programs, but not all of the many small
effects discussed above 
have been taken into account while the dominating source of 
uncertainty is still the statistical error.


\section{The Determination of $M_W$ in Charged Current Deep Inelatic 
Scattering}
\label{sec:ken}
\subsection{Introduction}
\label{Sub:MW_Intro}

Recently the ZEUS and H1 experiments at the HERA positron-proton
collider announced preliminary measurements of the single and double
differential cross sections for the CC reaction 
$e^+ P \rightarrow \bar{\nu}_e X$ \cite{Ref:H1Van, Ref:ZEUSVan}.
The measurements were made using data sets corresponding to a
luminosity of $\sim 40$~pb$^{-1}$ per experiment.
The CC data is now sufficiently precise to make it timely to
re-examine the extent to which the HERA experiments can contribute to
constraints on the electroweak sector of the SM.
In the following the statistical precision with which the parameters
$M_W$ or $G_\mu$ may be determined will be investigated.
Hence, it is the dependence of the CC cross section on
$G_\mu$ or $M_W$ which is of primary importance and small differences between
different approaches to the calculation of the radiative corrections
themselves will not be discussed.
There is no attempt in what follows to discuss the experimental
systematic uncertainties involved in measurements such as those
proposed.

\subsection{Sensitivity of CC DIS to $M_W$}

The Born cross-section for the reaction $e^+P\rightarrow \bar{\nu}_e X$
has been given in Eq.~\ref{eq:CCxsec}.
In this expression the coupling of the $W$-boson to the fermions is
contained in $G_\mu$ while the $W$-boson propagator contributes the
factor $[M_W^2/(M_W^2+Q^2)]^2$.

  The mass of the exchanged boson can be determined from the $Q^2$
dependence of the CC cross section since the propagator factor 
($[M_W^2/(M_W^2+Q^2)]^2$) depends explicitly on $M_W$.
This is the `propagator' mass mentioned in Sec.~\ref{sec:hubert} 
(scenario 1 with fixed $G_\mu$).
Both the ZEUS and H1 collaborations have performed such a
determination with the results \cite{Ref:H1Van, Ref:ZEUSVan}
\begin{eqnarray}
  M_W & = & 78.6^{+2.5}_{-2.4}\,{\rm [stat]}\,
                ^{+3.3}_{-3.0}\,{\rm [syst]}\, 
                ^{+1.5}_{-1.5}\,{\rm [PDF]}\,{\rm GeV \,\,\,\, (ZEUS),} \\
  M_W & = & 81.2^{+3.3}_{-3.3}\,{\rm [stat]}
                ^{+4.3}_{-4.3}\,{\rm [syst]}\,
                ^{+2.8}_{-2.8}\,{\rm [PDF]}\,{\rm GeV \,\,\,\, (H1).}  
\end{eqnarray}
The uncertainty on the value of $M_W$ arising from the 
PDFs is quoted separately above.
This error is determined by performing the fit
using a variety of PDFs (additionally, the error quoted by H1 
includes a 4\% error on the CC radiative corrections). 

  The data may also be used to determine $G_\mu$ or $M_W$ 
(assuming scenario 2).
The precision with which $G_\mu$ can be determined is directly related
to the precision with which the cross section is measured.
The size of the $e^+ P$ data set available to ZEUS and H1 yields
$\sim 1000$ CC events per experiment. Hence the
statistical uncertainty on such a direct determination of $G_\mu$ will
be $\sim 1.7$\%. 
Note that at HERA such a determination of $G_\mu$ will be
performed at an effective $Q^2$ in excess of $\sim 400$~GeV$^2$ and
hence will be complementary to the precise measurement of $G_\mu$
obtained from muon decay where the momentum scale is set by the mass
of the muon.

\begin{figure}[ht]
  \begin{center}
     \includegraphics*[width=\textwidth]{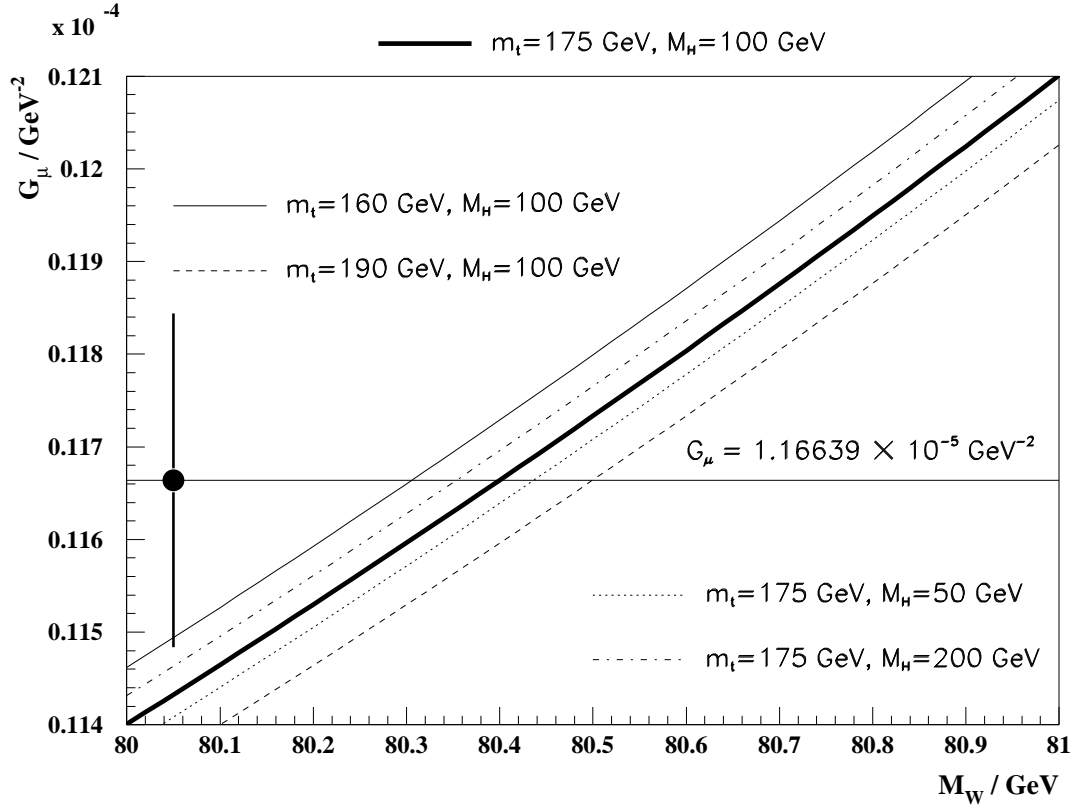}
  \end{center}
  \caption{
    The dependence of $G_\mu$ on $M_W$ embodied by the 
    $\mathcal{O}\left( \alpha \right)$ expression of equation
    \ref{rc6}.
    The thick solid line shows the result when $m_t$ is set to
    175~GeV and $M_H$ to 100~GeV. 
    The thin solid, dashed, dotted and dash dotted lines show the
    dependence of $G_\mu$ on $M_W$ for the values of $m_t$ and $M_H$
    indicated.
    The point with error bar shows the expected statistical
    uncertainty on $G_\mu$.}
  \label{Fig:GFvsMW}
\end{figure}

  At $\mathcal{O} \left( \alpha \right)$ in the SM $G_\mu$ depends upon
$M_W$ through Eq.~\ref{rc6}.
In the following we shall set $\alpha$, $M_Z$, and all fermion masses,
other than $m_t$ to the values quoted in reference \cite{Ref:PDG}.
As shown in Fig~\ref{Fig:GFvsMW}, $G_\mu$ depends strongly on $M_W$ 
and less strongly on the mass of the Higgs boson, $M_H$, or the mass
of the top quark, $m_t$.
Hence, within the context of the SM, the greatest sensitivity to $M_W$
may be obtained by combining the $M_W$ dependence of the propagator
term in the CC cross section with the $M_W$ dependence of $G_\mu$.
A naive propagation of the statistical error from a sample of 1000 CC
events leads to an error of $\sim 300$~MeV on the parameter $M_W$.
More detailed numerical estimates are given in the next section.
The figure shows the expected statistical error on a measurement of
$G_\mu$.
It can be seen that the variation of $G_\mu$ with $m_t$ or $M_H$ at
fixed $M_W$ is negligible compared to the expected statistical error.

\begin{figure}[htbp]
  \begin{center}
     \includegraphics*[height=.9\textheight]{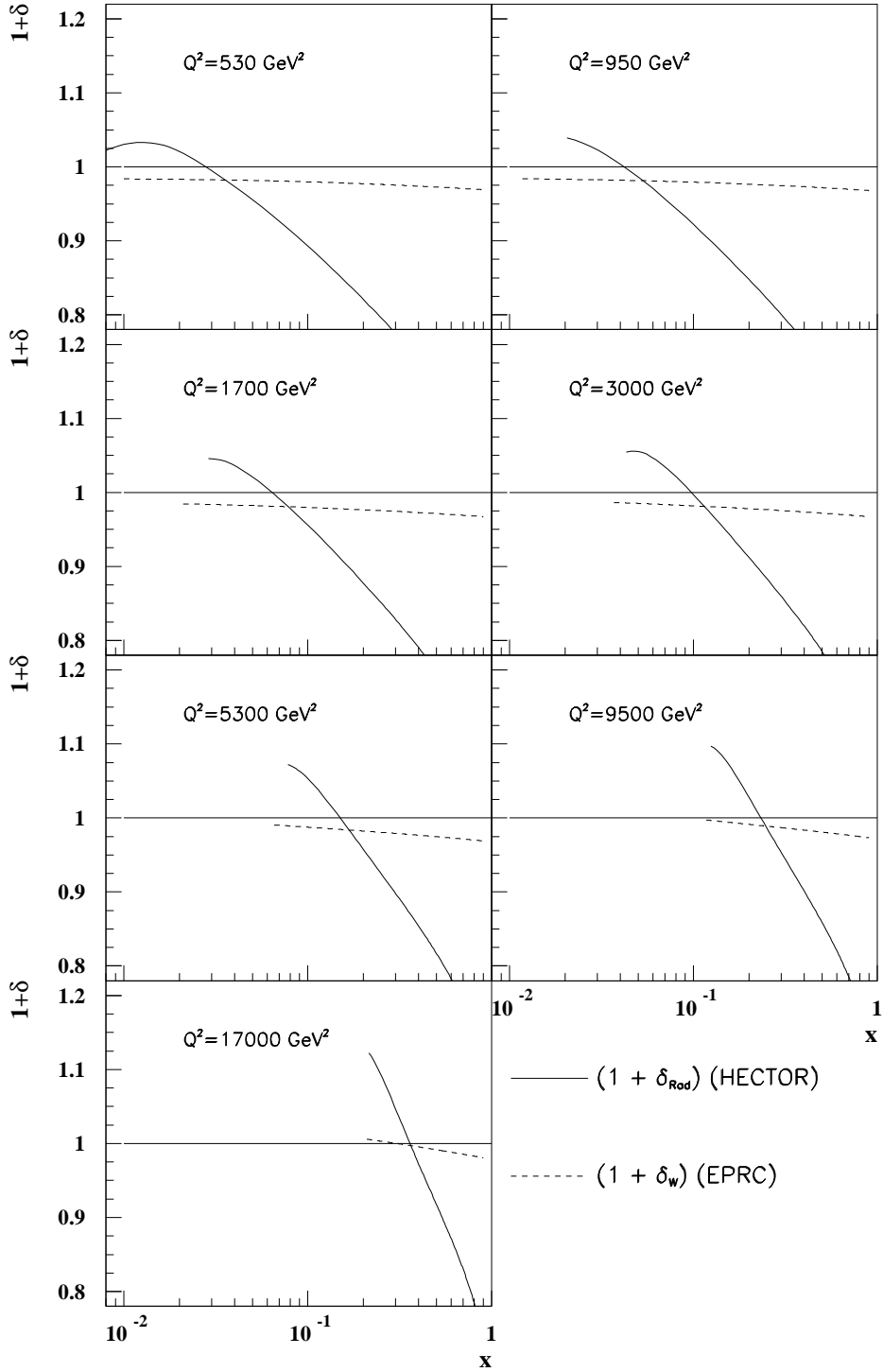}
  \end{center}
  \caption{
           Radiative corrections to charged current deep inelastic 
           scattering.
           The solid line shows the full radiative correction, 
           $\delta_{\rm Rad}$, calculated using the program HECTOR
           \protect{\cite{hector}}.
           The dashed line shows the electroweak contribution to the
           radiative corrections, $\delta_{\rm W}$, calculated using 
           the program EPRC \protect{\cite{hseprc}}.
           }
  \label{Fig:Delta}
\end{figure}

  Before proceeding to numerical estimates of the precision with which
$M_W$ or $G_\mu$ can be extracted from present CC data it is necessary
to consider the dependence of the radiative corrections to 
the Born cross section (Eq.~\ref{eq:CCxsec}) on $M_W$.
The radiative corrections to the Born cross section given in Eq.~\ref{rc5}
may be formally expressed in terms of a correction
factor, $\delta_{\rm Rad}$, defined by the equation
\begin{equation}
  {{d^2\sigma^{\rm CC}_{\rm Rad}}  \over {dx \, dQ^2}} =
  {{d^2\sigma^{\rm CC}_{\rm Born}} \over {dx \, dQ^2}} 
  \left[ 1 + \delta_{\rm Rad} \right].
\end{equation}
The full $\mathcal{O} \left( \alpha \right)$ radiative corrections may
be split into two pieces by writing
\begin{equation}
  1 + \delta_{\rm Rad} = \left( 1 + \delta_{\gamma}
                                  + \delta_{\rm W}  \right).
\end{equation}
A real photon may be radiated from one of the particles involved in
the hard scattering or appear as a virtual particle in a one-loop diagram.
Such corrections contribute to the term $\delta_{\gamma}$. 
All other $\mathcal{O} \left( \alpha \right)$ contributions
are included in the term $\delta_{\rm W}$.
The size of these corrections is shown in Fig~\ref{Fig:Delta}.
The full radiative correction, $\delta_{\rm Rad}$, was calculated using
the program HECTOR \cite{hector}, while $\delta_{\rm W}$ was
calculated using the program EPRC \cite{hseprc}.
It can be seen from Fig~\ref{Fig:Delta} that $\delta_{\rm Rad}$ can
be as large as $\pm 20 \%$ while $\delta_{\rm W}$ is never larger than 
$\sim 3 \%$.

\begin{figure}[htbp]
  \begin{center}
     \includegraphics*[height=.9\textheight]{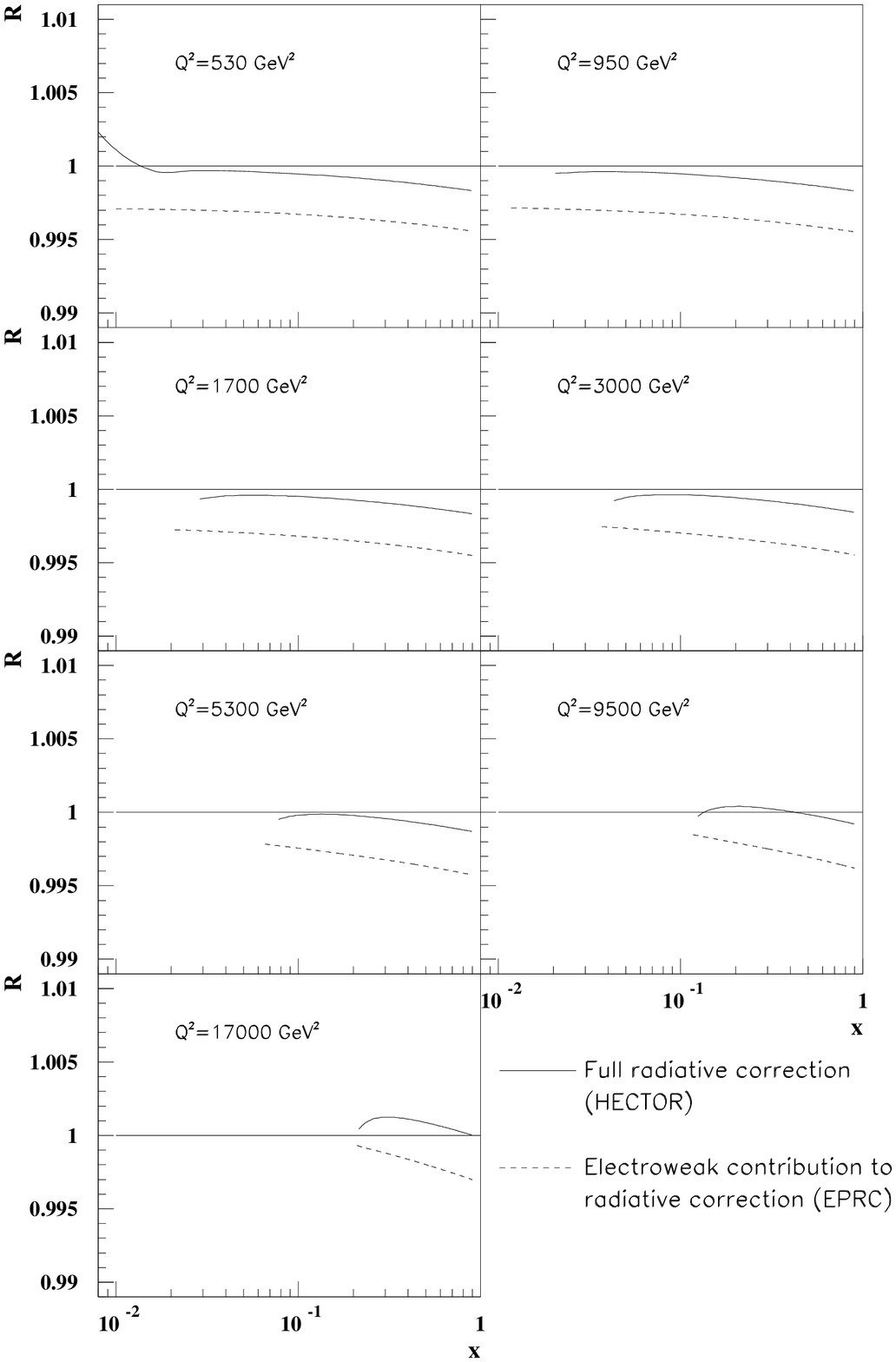}
  \end{center}
  \caption{
           The dependence of the radiative corrections to charged
           current deep inelastic scattering on the parameter $G_\mu$.
           The ratio $R$, defined in the text, shows the fractional
           change in the radiative correction as $G_\mu$ is varied from
           $1.116 \times 10^-5 {\rm~GeV}^{-2}$ to
           $1.226 \times 10^-5 {\rm~GeV}^{-2}$.
           The solid line shows the sensitivity of the full radiative
           correction (evaluated using HECTOR
           \protect{\cite{hector}}) to $G_\mu$.
           The dashed line
           shows the sensitivity of the electroweak contribution to the
           radiative corrections (calculated using the program EPRC 
           \protect{\cite{hseprc}}) to $G_\mu$.
           }
  \label{Fig:Rat}
\end{figure}

The radiative corrections to CC DIS depend on $G_\mu$; i.e. 
$\delta = \delta \left( G_\mu \right)$, where $\delta$ may be either
$\delta_{\rm Rad}$ or $\delta_{\rm W}$.
Thus for the determination of $G_\mu$ (or $M_W$) from the Born cross section
to be valid it is important that $\delta_{\rm Rad}$ should depend
only weakly on $G_\mu$.
The sensitivity of $\delta_{\rm Rad}$ and $\delta_{\rm W}$ to $G_\mu$
is shown in Fig~\ref{Fig:Rat} where the ratio
\begin{equation}
    R = \frac{1+\delta\left( G_\mu=1.226 \times 10^-5 {\rm~GeV}^{-2} \right)} 
             {1+\delta\left( G_\mu=1.166 \times 10^-5 {\rm~GeV}^{-2} \right)}
    \label{Eq:R}
\end{equation}
is plotted as a function of $x$ for several fixed values of $Q^2$.
The reference value of $G_\mu$ ($G_\mu = 1.166389 \times 10^{-5}$~GeV$^{-2}$)  
was taken from reference \cite{Ref:PDG}.
Given a statistical error of $\sim 1.7\%$ the value 
$G_\mu = 1.226 \times 10^{-5}$~GeV$^{-2}$ corresponds to raising $G_\mu$
by three standard deviations from the reference value.
Fig~\ref{Fig:Rat} shows that the sensitivity of $\delta_{\rm W}$ to
such a change in $G_\mu$ is small ($<1\%$).
The full radiative correction, $\delta_{\rm Rad}$, shows a still
weaker dependence on $G_\mu$ since $\delta_{\rm Rad}$ is dominated by
the photonic corrections contained in $\delta_{\gamma}$.

Given the weak dependence of the radiative corrections on $G_\mu$ it is
reasonable to estimate the precision with which $G_\mu$ (or $M_W$) may be
extracted by making a fit to the Born cross section.
In practice, once $G_\mu$ has been determined from such a fit, the
extraction of the experimental cross section would have to be iterated using
the new value of $G_\mu$ in order to verify that the results obtained
were stable against such modification.


\subsection{Numerical Estimates}
\label{Sub:MW_Numer}

  The program HERACLES \cite{Ref:HERACLES} was used to estimate the
statistical error corresponding to a luminosity of 40~pb$^{-1}$ of
$e^+P$ data (a total of 1020 CC events) distributed among eight bins
in $Q^2$. 
The bins were equally spaced in $\log_{10} Q^2$ for $Q^2$ in the range
$400 < Q^2 < 40000$~GeV$^2$. Eq~\ref{eq:CCxsec} was then used to
evaluate $d\sigma^{\rm CC}_{\rm Born} / d Q^2$.
The CTEQ4D PDFs were used through out.

  An estimate of the precision with which $G_\mu$ may be determined was 
obtained as follows.
The value of $M_W$ was fixed at $M_W = 80.41$~GeV and a $\chi^2$ fit 
of the differential cross section, $d\sigma(G_\mu) / dQ^2$, performed.
This is scenario 1 with $M_W$ fixed, as described in Sec.~\ref{sec:hubert}.
The fit yielded an error on 
$G_\mu$ of $\pm 0.018 \times 10^{-5}\,{\rm GeV^{-2}}$ 
which corresponds to $\pm 1.5\%$.
A slightly improved error may be obtained by considering scenario 2 which
combines the sensitivity to the propagator and to the normalization into one
common parameter, using Eq.~\ref{rc6}. The error on $G_\mu$ is then 
$\pm 0.016 \times 10^{-5}\,{\rm  GeV^{-2}}$
($\pm 1.4\%$). 
This last result may be re-expressed to give an estimate of the error on the
value of $M_W$, using Eq.~\ref{rc9}, as suggested in Sec.~\ref{sec:hubert}.
Such a fit was performed and yielded a statistical error on the
parameter $M_W$ of $\pm 0.24$~GeV.
These results agree well with similar fits reported in references
\cite{EWheraB} and \cite{Ref:Elsen}.

\section{Summary and Outlook}
\label{sec:summary}

We now consider what we have learnt and what we can learn given various
running conditions: $e^+/e^-$ beams, polarization of the beam, beam energies. 

We have seen that HERA data has been used to measure the `propagator' mass
$M_W$ and that the precision on such a measurement may be improved by using
information on the normalization as well as the shape of the cross-section.
Future improvements in luminosity ($\sim 1000pb^{-1}$) 
may allow us to achieve an error of 
$\pm55$MeV on $M_W$~\cite{Ref:Elsen}. Sensitivity is greatest with electrons 
rather than positrons for both NC and CC processes and NC data is most useful 
with left handed 
polarization of the lepton beam (which enhances the $\gamma/Z$ terms, see
 Eqs.~\ref{eq:NCxsec}-~\ref{eq:bi}). Indeed 70\% polarization is worth a 
factor of 4 in luminosity. 


One may also be able to measure the weak neutral current 
couplings of quarks, $v_u$, $a_u$, $v_d$, $a_d$ 
(see Eqs.~\ref{eq:NCxsec}-~\ref{eq:bi}),
as explored in Ref.~\cite{cashmore}. With unpolarized $e^+$ and
$e^-$ beams ratios of NC and CC cross-sections can be used to determine
$a_u$, whereas polarized beams allow the extraction of $v_u$.
Measurements of NC cross-sections alone may be used in a fit to determine
all four couplings. Polarization of the beams is essential to achieve 
reasonable precision. One needs $\simeq 250pb^{-1}$ luminosity for
each of the four lepton beam charge/polarization combinations to make 
measurements with a precision of $\sim 10\%$.

We have seen that HERA CC data can improve
our knowledge of the $u$ and $d$ valence distributions 
(Sec.~\ref{sec:bodek}). It seems that knowledge of the
$d$ quark is the most crucial since the $u$ quark is already much better
known, not being subject to uncertainties due to deuteron binding corrections.
To improve our knowledge we will require further positron beam running.
Since we would like to explore the valence distributions at the highest 
possible values of $x$ it may also be worth lowering the proton beam energy
to achieve higher $x$ values at the same $Q^2$ ($x=Q^2/(ys)$). This would have
two further advantages: it would allow substantial overlap with fixed target 
data at lower $Q^2$ and thus constrain 
normalizations, and
it would allow a model independent  measurement of the longitudinal structure 
function $F_L$~\cite{AMCS}.

Can we extract other interesting structure functions or PDFs with future
high luminosity, high $Q^2$, NC and CC data? This has been explored 
in~\cite {ruckl}. Polarization is not explicitly useful for this purpose
so we will assume $P=0$ in the formulae (though they can be generalized to
any value of $P$). If we express Eqs.~\ref{eq:empCC} and~\ref{eq:eppCC} 
in terms of reduced cross-sections we obtain
\begin{equation}
 \tilde{\sigma}_{CC}(e^-p) = \left[x(u+c) +(1-y)^2 x(\bar d+\bar s)\right]
\label{eq:charm}
\end{equation}
\begin{equation}
 \tilde{\sigma}_{CC}(e^+p) = \left[x(\bar u+\bar c) +(1-y)^2 x(d+s)\right].
\label{eq:strange}
\end{equation}
Then we can define the sum and difference of these reduced cross-sections
$\tilde{\sigma}_{\pm}(CC)=\tilde{\sigma}_{CC}(e^-p) \pm
\tilde{\sigma}_{CC}(e+p)$, so that
\begin{equation}
 \tilde{\sigma}_+(CC) = xU + (1-y)^2 xD
\end{equation}
\begin{equation}
 \tilde{\sigma}_-(CC) = xu_v - (1-y)^2 xd_v
\end{equation}
where $U$ stands for the sum of the $u$-type quarks and antiquarks 
($u$,$\bar u$, $c$, $\bar c$), $D$ for the sum of the $d$-type
($d$, $\bar d$, $s$, $\bar s$) and $u_v$ and $d_v$ stand for the valence
quark densities.
Similarly the sum and difference of the NC $e^+p$ and $e^-p$ 
reduced cross-sections give
\begin{equation}
 \tilde{\sigma}_+(NC) = A_u xU + A_d xD = F_2
\end{equation}
\begin{equation}
 \tilde{\sigma}_-(NC) = B_u xu_v +  B_d xd_v = xF_3
\end{equation}
where $A_u$, $B_u$ are the same for all $u$-type quarks and antiquarks and
$A_d$, $B_d$ are the same for all $d$-type quarks and antiquarks.

Thus from the CC cross-sections we can mainly gain information on the valence 
distributions at high $x$, whereas from the NC cross-sections
 we may measure $xF_3$ for the first time.
We may also consider making combinations of all four cross-sections
$\tilde{\sigma}_{+/-}(NC/CC)$ in order to extract $U$, $D$, $u_v$, $d_v$
for the complete $x$ range. A detailed analysis of this possibility indicates 
that $200pb^{-1}$ per $e^+/e^-$ beam will be necessary to make measurements
of $\sim 10\%$ accuracy~\cite{ruckl}.
Finally, can HERA shed light on the charm and strange components of the sea?
Eqs.~\ref{eq:charm}, ~\ref{eq:strange} indicate how the CC processes may be 
studied at low $x$ and high and low $y$ to look for evidence of the 
appropriate sign of $D$ or $K$ hadrons in the final state.

In summary it seems that running conditions with an equal partition between
$e^+/e-$ and left/right handed polarization will give the most flexibilty
for future physics, and that running at a reduced proton beam energy is of 
interest to physics at high $Q^2$ as well as at low $Q^2$.

\ack
The high $Q^2$ working group would like to acknowledge the contribution of all
particpants who attended the sessions, with particular thanks to C. Cormack.
     

\end{document}